\documentclass[a4paper,12pt]{article}
\usepackage{latexsym,amssymb}
\usepackage{amsmath,amssymb,amsxtra,amscd,amsthm,amsfonts}

\def\ma{1\!{\rm l}}

\def\a{\alpha}
\def\b{\beta}

\def\e{\epsilon}

\def\p{\partial}

\def\e{\epsilon}

\def\Tr{{\rm Tr}}
\def\P1{{\bf P}^1}
\def\ln{{\rm ln}}
\def\0{{\nonumber}}
\def\be{\begin{equation}}
\def\ee{\end{equation}}
\newcommand{\mat}[2][ccccccccccccccccccccccccccccccccccccccccc]{\left(
\begin{array}{#1}
#2\\
\end{array}
\right)}

\newcommand{\br}[2][lllllllllllllllllllllllllllllllll]{\left\lbrace
\begin{array}{#1}
#2\\
\end{array}
\right.}

\begin{document}
\begin{titlepage}
\flushright{\hfill SISSA/48/2008/EP} 
\flushright{\hfill DISTA-2008}

\vspace{2.5cm}

\centerline{\bf\LARGE{Exploring Pure Spinor String Theory }}
\centerline{\bf\LARGE{ on $AdS_4\times \mathbb{CP}^3$} }
\vspace{3.0cm}

\centerline{\large Giulio Bonelli$^{~a}$, Pietro Antonio Grassi$^{~b}$, and Houman Safaai$^{~a}$}
\vskip .7cm
\centerline{$a)$~International School of Advanced Studies (SISSA)} 
\centerline{ via Beirut 2-4, 34014 Trieste, Italy}
\centerline{INFN - Sezione di Trieste}

\vskip .2cm
\centerline{$b)$~DISTA, Universit\`a del Piemonte Orientale,}  
\centerline{via Bellini 25/G,  Alessandria, 15100, Italy}
\centerline{INFN - Sezione di Torino, Gruppo Collegato di Alessandria}

\vspace{3.5cm}

\begin{abstract}
In this paper we formulate the pure spinor superstring theory on $AdS_4\times \mathbb{CP}^3$.
By recasting the pure spinor action
as a topological A-model on the fermionic supercoset $Osp(6|4)/SO(6)\times Sp(4)$
plus a BRST exact term,
we prove the exactness of the $\sigma$-model.
We then give a gauged linear $\sigma$-model which reduces to the superstring in the limit of large volume
and we study its branch geometry in different phases.
Moreover, we discuss possible D-brane boundary conditions and the principal chiral model for the 
fermionic supercoset.
\end{abstract}
\end{titlepage}

\vfill
\eject

\tableofcontents

\section{Introduction}

The pure spinor formulation of superstrings \cite{purespinors} is a powerfull method to tame 
superconformal exactness, the string loop expansion and RR-background $\sigma$-model couplings
at the same time.

In particular, let us remark that, although conjectured because of maximal supersymmetry,
the quantum exactness of type IIB string on $AdS_5\times S^5$ was proved in \cite{quantumex}
by making use of the pure spinor formulation.

Recently, the M-theory analog $AdS_4\times S^7$ is  receiving large attentions \cite{AharonyUG,disastro}
because of its conjectured duality with the Bagger-Lambert-Gustavsson 
\cite{BLG} theory of multiple M2-branes.
As a superstring theory, because of the circle fibration 
$$
\begin{matrix}
S^1 & \hookrightarrow & S^7 \\
{}  &   {}  &  \downarrow \\
{}  &   {}  & {\mathbb P}^3
\end{matrix}
$$
this is represented as type IIA superstring on $AdS_4\times \mathbb{CP}^3$ with appropriate RR-fluxes
turned on. This superstring background is undergoing an intense study
\cite{frolov,stefanski,antonio,AdS-4strings}, but its exact superconformal invariance has not been established so far.
This is one motivation\footnote{For further motivations and results on the $AdS_4\times \mathbb{CP}^3$ from a supergravity point of view, 
see also \cite{Tomasiello}.}
to study the pure spinor formulation of superstrings on $AdS_4\times \mathbb{CP}^3$ and
in this paper we actually prove its superconformal exactness.
This is done by writing the pure spinor action as a manifestly superconformal term plus a BRST trivial term.
Under the assumption that only mild non-locality arises in the BRST trivial term, we establish superconformal exactness.

Let us point out some further considerations about the system for $AdS_4 \times \mathbb{CP}^3$. In the present 
background there is a RR flux balancing the spacetime curvature which can be tuned to reach opposite limits: the strong coupling 
limit where the RR fields become dominant and the weak coupling limit where the supergravity approximation is valid. 
In the case of $AdS_5 \times S^5$, the two limits where covered by the same theory, namely $N=4, d=4$ SYM and the 
two limits of string theory where represented by the opposite limits of the gauge theory side. In that perspective, the supergravity computations 
in $AdS_5 \times S^5$ background leads to strong coupling correlation functions, whereas the strong coupling limit in string theory (where the 
supergravity approximation is no longer a good one) corresponds to perturbative SYM at weak coupling. 

Recently, in \cite{Berkovits1} and in \cite{BV}, it has been conjectured that this limit can be achieved by constructing the pure spinor sigma model 
on the coset $PSU(2,2|4)/ SO(1,4) \times SO(5)$, by taking the limit where the coupling constant goes to infinity and, finally, by adding a BRST trivial term, one can recast the $\sigma$-model into a non-linear sigma model which is a topological A-model that can be proved to be conformal to all orders (given the fact that 
the sigma model is based on a symmetric coset and the supergroup $PSU(2,2|4)$ is a super-Calabi-Yau). 
This construction has been tested in \cite{BS} where, by using a mirror symmetry argument, this program was realized 
for circular planar 1/2-BPS Wilson loops. 
See also \cite{brenno} for further developments.
Actually, in 
\cite{BerkovitsQC}, the relation between perturbative SYM $N=4$ and the topological sigma model has been further developed. 
There, it has been noticed that given a suitable measure for integrating the pure spinors in tree level amplitudes and 
identifying the vertex operators of the topological sigma model 
with the states of the fundamental representation of $PSU(2,2|4)$ (known as singleton), one can define some correlation functions with the properties 
of tree level SYM amplitudes. However, several checks and computations should be performed to test this new idea. Nevertheless, it has been shown that 
the topological $\sigma$-model can be viewed as a gauged linear $\sigma$-model of the G/G type with a suitable gauge fixing. 

On another side, we would like to perform the same analysis with another gauged linear sigma model. We consider the gauged sigma model 
with the supergroup $Osp(6|4)$. It has been already shown that this model leads to a pure spinor string theory model and its action has been constructed. 
In the present work we construct the gauged sigma model by gauge fixing the gauge symmetries of $Osp(6|4)$. Again, there are two limits. One limit 
is the weak coupling limit where the supergravity approximation is valid (see for example \cite{AharonyUG}) and this leads to multiple M2 brane interpretation 
of its dual theory (based on Bagger-Lambert-Gustavsson theory). This model is a superconformal Chern-Simons theory in $d=3$ with $N=6$. Notice that 
in this model the kinetic term of the gauge field is neglected since it has a dimensionful coupling, corresponding to the inverse of 
radius of $AdS$ which, in the limit of small RR flux $e$, tends to zero. 
The only remaining gauge field dynamics is described by a Chern-Simons model. However, being supersymmetric, it has some auxiliary fields which have to be integrated leading to the potential for the matter fields. So, the limit of strong coupling is the limit explored in the weak 
coupling limit of string theory. On the other  side, by considering the opposite limit, namely for large RR flux $e$, one should see the opposite 
limit of perturbative SYM d=3 model (with the kinetic term). 
This model has been constructed in \cite{FabbriAY}. So, we expect that in this limit the theory is no longer superconformal and contains some 
dynamical gauge field. The perturbative computation, which can be performed using the singleton conjecture of \cite{BerkovitsQC} 
should not correspond to 
the strong coupling limit of N=6 d=3 model, but to the weak coupling limit of $N=6, d=3, SYM$ (the symmetry of this model could be maybe enhanced 
to $N=8$).  


The plan of the paper is as follows.
In the next section we implement and solve the pure spinor constraints on $AdS_4\times \mathbb{CP}^3$
and therefore we formulate the pure spinor superstring theory on this background.
In section 3 we study the A-model topological string on the fermionic supercoset $Osp(6|4)/SO(6)\times Sp(4)$
and we establish its superconformal invariance.
In section 4 we show that the pure spinor action for the superstring on 
$AdS_4\times \mathbb{CP}^3$
can be recasted up to an additive BRST exact term as the topological A-model
on the fermionic supercoset $Osp(6|4)/SO(6)\times Sp(4)$.
Actually, this is obtained as a particular case of a more general construction for supercosets
admitting a compatible $\mathbb Z_4$ grading.
In section 5 we then give a gauged linear $\sigma$-model which reduces to the topological string in the limit of large 
Fayet-Illiopoulos coupling and we study its Coulomb branch geometry.
In section 6 we formulate a principal chiral model of $G/G$ type which upon gauge fixing 
reduces to the A-model on the supercoset.
Moreover, in Section 7 we discuss possible D-brane boundary conditions and discuss their geometric structure.
We indicate further directions to explore in Section 8 where we collect some open issues too.

\section{Pure spinor superstring in $AdS_4\times \mathbb{CP}^3$ background}

As it was shown in \cite{antonio,frolov,stefanski}, the $AdS_4\times \mathbb{CP}^3$ background can be derived from a supercoset element $g\in\frac{Osp(6|4)}{U(3)\times SO(1,3)}$. Its Maurer-Cartan left invariant 1-form can be expanded into the generators of $Osp(6|4)$ as follows
\begin{equation}
 J=J^a \gamma_a+J_{IJ} T^{IJ}+J^{IJ} T_{IJ}+H^{ab}\gamma_{ab}+H_I^{\;J} T_J^{\;I}+J_I^{\;\alpha} Q^{\;I}_\alpha+J_I^{\;\dot\alpha} Q^{\;I}_{\dot\alpha}+J^{I\alpha} Q_{I\alpha}+J^{I\dot\alpha} Q_{I\dot\alpha}\,,
\end{equation}



where $(T_{IJ},T^{IJ},T_J^{\;I})$ are the generators of $SO(6)$, $T_{[AB]}$ with $A,B=1\ldots 6$ decomposes according to irreducible reperesentations of $U(3)$ as it will be explained later, and $T_J^{\;I}$ are the generators of $U(3)$. Then, $J_{IJ}$ and $J^{IJ}$ are the Maurer-Cartan forms associated to the generators of the coset $\frac{SU(4)}{U(3)}$ and $H_I^{\;J}$ are the corresponding spin connections of the coset. Similarly, $(\gamma_a,\gamma_{ab})$ with $a,b=1\ldots 4$ are the generators of the anti de Sitter group $SO(2,3)$ which as is shown in \cite{antonio} they all turn out to be given by real symplectic matrices and $\gamma_{ab}$ are the generators of the Lorentz group $SO(1,3)$. The matrices $Q_I^{\;\alpha},Q_I^{\;\dot\alpha},Q^{\;I}_\alpha$ and $Q^{\;I}_{\dot\alpha}$ are the 24 fermionic generators where we split the symplectic indices $x=1\ldots 4$ into $SO(1,3)$ spinorial indices $\alpha,\dot\alpha=1,2$. The Maurer-Cartan 1-forms of the symplectic group $Sp(4,\mathbb{R})$ are related to the Maurer-Cartan of $SO(2,3)$ with the relation $J^{xy}=J^a \gamma_a^{xy}+H^{ab} \gamma_{ab}^{xy}$. The fermionic 1-forms $J_A^x$ are real and transform in the fundamental 4-dimensional representation of $\mathfrak{sp}(4,\mathbb{R})$ and in the fundamental 6-dimensional representation of $\mathfrak{so}(6)$ with the symplectic invariant antisymmetric metric  $\epsilon_{xy}=i\sigma_1\otimes \ma$.

Notice that $\eta^{ab}$ is the invariant metric on $AdS_4$ and $g_{I\bar J}$ is the $U(3)$ invariant metric on $\mathbb{P}^3$ and we denote by $k_{I\bar J}$ as the K\"ahler form on $\mathbb{P}^3$.
The index $I$ can be raised and lowered with the inverse metric $g^{\bar I J}$ as $J^{\bar I \bar J}=g^{\bar I K}g^{\bar J L} J_{KL}$ which is independent of $J^{IJ}$, similarly we can make $J_{\bar I \bar J}$ out of $J_{IJ}$. 

The $\mathfrak{osp}(6|4)$ algebra $\mathcal{H}$ admits a $\mathbb{Z}_4$ grading with decomposition $\mathcal{H}=\sum_{i=0}^3 \mathcal{H}_i$ as follows\footnote{In the paper, also the notation $\hat J$ will be used to denote the currents of the subset $\mathcal{H}_3$.}
\begin{eqnarray}\label{Z4}
{\mathcal H}_0 &=& \Big\{ H_{\alpha\beta},H_{\dot\alpha\dot\beta},H_{I}^{\;J} \Big\}\,,  \hspace{0.93cm}
{\mathcal H}_1 = \Big\{ J^{\alpha  I},J^{\dot\alpha \bar I} \Big\}\,, \nonumber \\
{\mathcal H}_2 &=& \Big\{ J_{\alpha\dot\alpha}, J_{IJ}, J^{IJ} \Big\}\,, ~~~~~~~\hspace{1.05cm}
{\mathcal H}_3 = \Big\{ J^{\,\,\alpha}_{I},J^{\,\,\dot\alpha}_{\bar I } \Big\}\,.
\end{eqnarray} 
satisfying
\begin{equation}
 \left[ {\mathcal H}_m,{\mathcal H}_n\right]\subset {\mathcal H}_{m+n\;(\text{mod }4)}
\end{equation}

We can check that the bilinear metric is also ${\mathbb Z}_4$ invariant. Recall that the invariant supermetric for $Osp(6|4)$ is given by 
\begin{eqnarray}\label{bil1}
{\rm Str}( T_{AB} T_{CD} ) &=& \delta_{A C} \delta_{D B} - \delta_{A D} \delta_{C B} \,, 
\\ \nonumber
{\rm Str}(T_{xy}\, T_{zt}) &=& \epsilon_{x z} \epsilon_{t y} + \epsilon_{x t} \epsilon_{z y} \,, 
\\ \nonumber
{\rm Str}(T_{x}\, T_{y}) &=& \epsilon_{x y} \,, 
\\ \nonumber
{\rm Str}( Q^x_A Q^y_B ) &=& \delta_{AB} \epsilon^{xy}\,.
\end{eqnarray}
where $T_{AB}$ and $T_{xy}$ are the generators of the bosonic subgroups $SO(6)$ and $Sp(4,\mathbb{R})$, and 
$Q^x_A$ are the fermionic generators of the supergroup. It is convenient to adopt a complex basis for the 
generators of $SO(6)$ and we define $T_{AB} = U^{IJ}_{AB} T_{IJ} +  U^{I}_{J, AB} T_{I}^{~J}+  U_{IJ, AB} T^{IJ}$ where 
$U^{IJ}_{AB}, U^{I}_{J, AB}, U_{IJ, AB}$ are the Clebsh-Gordon matrices mapping from ${\underline{15}}$  of $SO(6)$ 
to the representations ${\underline 3}(-1)$, ${\underline 8}(0)$, ${\underline 3}^*(+1)$ of $U(3)$, respectively.
In the same way, we decompose the fermionic generators into $Q^x_I$ and $Q^{x I}$ of ${\underline 3}(-1)$ and 
 ${\underline 3}^*(1)$, respectively. 
The metric becomes 
\begin{eqnarray}\label{bil2}
{\rm Str}( T_{IJ} T^{KL} ) &=& \delta_{I}^{~K} \delta_{J}^{~L} - \delta_{J}^{~K} \delta_{I}^{~L}\,, 
\\ \nonumber
{\rm Str}( T_{I}^{~J} T_K^{~L} ) &=& \delta_{I}^{~L} \delta^{~J}_{K}\,,~~~~~
\\ \nonumber
{\rm Str}( Q^x_I Q^{y J} ) &=& \delta_{I}^J \epsilon^{xy}\,.
\end{eqnarray}
while the other traces vanish. Which all these mean that the bilinear metric is ${\mathbb Z}_4$ invariant, satisfying
\begin{equation}
 <{\mathcal H}_m,{\mathcal H}_n>=\text{Str}({\mathcal H}_m{\mathcal H}_n)=0,\;\;\text{unless }m+n=0\;\text{mod }4
\end{equation}

Using this ${\mathbb Z}_4$ automorphism, it was shown that the pure spinor sigma model action can be decomposed in the following way
\begin{equation}
 S=S_{GS}+S_{GF}+S_{\text{ghost}} \,,
\end{equation}
where $S_{GS}$ is the Green-Schwarz action was shown in  \cite{antonio,frolov,stefanski} to exhibit the usual quadratic form after using the important feature of the possibility of writing the Wess-Zumino term as a total derivative in this background
\begin{equation}
 S_{GS}=R^2\displaystyle\int d^2 z  \text{Str}\left[ \frac{1}{2}J_{2}{\bar J}_{2}+\frac{1}{4}\left( J _{1}{\bar J} _{3}-{ J} _{3}{\bar J} _{1}\right) \right]\,,
\end{equation}
where $J_{i}=J|_{\mathcal{H}_i}$ are the projections of the MC left invariant currents into different subclasses according to $\mathbb{Z}_4$ automorphism as it was given in (\ref{Z4}). The action can be written in terms of the left-invariant supercurrents of the coset in the following form 
\begin{equation}
 S_{GS}=R^2\displaystyle\int d^2z \left[ \epsilon_{xy} J^x \bar{J}^y+\frac{1}{2} J_{IJ} \bar{J}^{IJ} + \frac{1}{4}\left( J_{\alpha I}{\bar J}^{\alpha I}+J_{\dot\alpha\bar I}{\bar J}^{\dot\alpha\bar I}-J_{\alpha \bar I}{\bar J}^{\alpha \bar I}-J_{\dot\alpha I}{\bar J}^{\dot\alpha I} \right)\right]\,.
\end{equation}

To this, one has to add a term which breaks $\kappa-$symmetry and adds kinetic terms for the target-space fermions and the coupling to the RR flux. This gauge fixing action $S_{GF}$ was shown to be given by \cite{antonio}
\begin{equation}
 S_{GF}=
R^2\displaystyle\int d^2 z \left( J_{\alpha \bar I}{\bar J}^{\alpha \bar I}+J_{\dot\alpha I}{\bar J}^{\dot\alpha I} \right)\,,
\end{equation}
which gives
\begin{equation}
 S_{GS}+S_{GF}=R^2\displaystyle\int d^2z \left[ \epsilon_{xy} J^x \bar{J}^y+\frac{1}{2} J_{IJ} \bar{J}^{IJ} + \frac{1}{4}\left( J_{\alpha I}{\bar J}^{\alpha I}+J_{\dot\alpha\bar I}{\bar J}^{\dot\alpha\bar I}\right)
+\frac{3}{4} \left( J_{\alpha \bar I}{\bar J}^{\alpha \bar I}+J_{\dot\alpha I}{\bar J}^{\dot\alpha I} 
\right)\right]\,.
\end{equation}

In order to write the pure spinor ghost part of the action, we introduce the pure spinors $(\lambda_I^{\,\,\alpha},\lambda_{\bar I}^{\,\,\dot\alpha})$, $({\hat\lambda}^{\alpha I},{\hat\lambda}^{\dot\alpha\bar  I})$ and their conjugate momenta $(w_{\alpha}^{\,\, I},w^{\,\,\bar I}_{\dot\alpha})$, 
$({\hat w}_{\alpha I},{\hat w}_{\dot\alpha\bar I})$, belonging to the ${\mathcal{H}}_1$ and ${\mathcal{H}}_3$  respectively. The pure spinor constraints can be written as follows
\begin{eqnarray}\label{pures}
\br{\lambda_I^\alpha \lambda^{\dot\alpha I}=0\\\lambda_I^\alpha \epsilon_{\alpha\beta}\lambda_J^{\beta}=0\\\lambda^{\dot\alpha I}\epsilon_{\dot\alpha\dot\beta}\lambda^{\dot\beta J}=0}
,\hspace{2cm}
\br{{\hat\lambda}^{\alpha I} {\hat\lambda}^{\dot\alpha}_I=0\\{\hat\lambda}^{I\alpha} \epsilon_{\alpha\beta}{\hat\lambda}^{\beta J}=0\\{\hat\lambda}^{\dot\alpha}_I \epsilon_{\dot\alpha\dot\beta}{\hat\lambda}^{\dot\beta}_J=0}
\end{eqnarray}
to solve this constraint, we can use the following ansatz 
\begin{eqnarray}
 \lambda^{\,\,\alpha}_I&=&\lambda^\alpha u_I,\;\;\;\;\;\;\;\lambda^{\dot\alpha I}=\lambda^{\dot\alpha}v^I\,,
\\ \nonumber
 {\hat\lambda}^{\alpha I}&=&{\hat\lambda}^\alpha {\hat u}^I,\;\;\;\;\;\;\;{\hat\lambda}^{\,\,\dot\alpha}_I={\hat\lambda}^{\dot\alpha}{\hat v}_I\,,
\end{eqnarray}
subject to the following gauge transformations
\begin{eqnarray}
 \lambda^\alpha\rightarrow \frac{1}{\rho}\lambda^\alpha,\;\;\;\;\; \lambda^{\dot\alpha}\rightarrow \frac{1}{\sigma} \lambda^{\dot\alpha},\;\;\;\;\; u_I\rightarrow \rho u_I,\;\;\;\;\; v^I\rightarrow \sigma v^I\,,
\\ \nonumber
 {\hat\lambda}^\alpha\rightarrow \frac{1}{\hat\rho}{\hat\lambda}^\alpha,\;\;\;\;\; {\hat\lambda}^{\dot\alpha}\rightarrow \frac{1}{\hat\sigma} {\hat\lambda}^{\dot\alpha},\;\;\;\;\; {\hat u}^I\rightarrow \hat\rho {\hat u}^I,\;\;\;\;\; {\hat v}_I\rightarrow \hat\sigma {\hat v}_I\,,
\end{eqnarray}
where $\rho,\sigma,\hat\rho,\hat\sigma\in\mathbb{C}^*$.

Inserting these factorization into (\ref{pures}), we arrive to the following constraints
\begin{eqnarray}
 u_Iv^I=0,\;\;\;\;\;\;\; {\hat v}_I{\hat u}^I=0\,.
\end{eqnarray}

So, the counting of the degrees of freedom gives $2\times(2+3-1)-1=7$ complex for $\lambda$ and the same for $\hat\lambda$. The geometry of the pure spinor space can be easily described. Using the gauge symmetries $\rho$ and $\sigma$ we can fix the norm of $u_I$ and $v^I$ as such 
$u_I \bar u^I = 1$ and $v^I \bar v_I = 1$. Then, together the constraint $u_I v^I =0$, the 
matrix $(u_I, \bar v_I, \epsilon_{IJK} \bar u^J v^K)$ is an $SU(3)$ matrix. In addition, using the 
remaining phases of the gauge symmetries $\rho$ and $\sigma$, we see that the variables $u_I$ and 
$v^I$ parametrize the space $SU(3)/ U(1) \times U(1)$ which is the space of the harmonic variables of the $N=3$ harmonic superspace (it is also known as the flag manifold $F(1,2,3)$.\footnote{Another way to solve the constraints (\ref{pures}) is decomposing the pure spinor into $\lambda^\a_I = ( \lambda^\a_a, \lambda^\a)$ and $\lambda^{\dot\a I} = (\lambda^{\dot \a a}, \lambda^{\dot \a})$ where $a=1,2$. It is easy to show that the pure spinor constrains become 
$\lambda^\a_a \lambda^{\dot\a a} + \lambda^\a \lambda^{\dot\a} =0$, $\det(\lambda^\a_a) =0$, $\det(\lambda^{\dot\a}_a) =0$, $\lambda^\a_a \e_{\a\b} \lambda^\b=0$ and $\lambda^{\dot\a}_a \e_{\dot\a\dot\b} \lambda^{\dot\b}=0$. The first set of constraints implies that we can solve 3 parameters in terms of the rest and we get a consistency condition $\det(\lambda^\a_a) \det(\lambda^{\dot\a}_a) =0$. This is solved by imposing the second and the third 
conditions. The latter also imply the existence of a solution for the forth and for the fifth constraints. Again the counting of the parameters gives 7 complex numbers.} 

The pure spinor constraints are first class constraints and they commute with the Hamiltonian, therefore 
they generat the gauge symmetries on the antighost fields $w$'s. In particular if we denote by 
$\eta_{\a\dot\a}, \eta^{IJ}, \eta_{IJ}$ and 
by $\kappa_{\a\dot\a}, \kappa^{IJ}, \kappa_{IJ}$ the infinitesimal parameters of the gauge symmetries 
we have that 
\begin{eqnarray}\label{gautra}
&&\delta w_\a^I = \eta_{\a\dot\a} \lambda^{\dot\a I} + 2 \eta^{IJ} \e_{\a\b} \, \lambda^\b_J\,, \hspace{2cm}\delta w_{\dot \a I} = \eta_{\a\dot\a} \lambda^{\a}_I + 2 \eta_{IJ} \e_{\dot\a\dot\b} \, \lambda^{\b J}\,, \nonumber \\
&&\delta \hat w_{\a I} = \kappa_{\a\dot\a} \hat\lambda^{\dot\a}_I + 
2 \kappa^{IJ} \e_{\a\b} \, \hat\lambda^{\b J}\,, \hspace{2cm}
\delta \hat w_{\dot \a}^I = \kappa_{\a\dot\a} 
\hat\lambda^{\a I} + 2 \eta^{IJ} \e_{\dot\a\dot\b} \, \hat\lambda^{\b}_I\,, 
\end{eqnarray}

We can also introduce the pure spinor Lorentz generators $(N=-\{w,\lambda\},\hat N=-\{\hat w,\hat\lambda\})\in {\mathcal{H}}_0$, which are needed in the action and determine the couplings between the pure spinor fields and matter fields, as follows
\begin{eqnarray}\label{lollo}
N_{\alpha\beta} &=& w^I_{(\alpha} \lambda_{\beta)I},\;\;\;\;\;\;\; {\hat N}_{\alpha\beta}= w_{I (\alpha} \lambda_{\beta)}^I\,,
\\ \nonumber
{ N}_{\dot\alpha\dot\beta} &=& { w}_{(\dot\alpha I} {\lambda}_{\dot\beta)}^I,\;\;\;\;\;\;\; {\hat N}_{\dot\alpha\dot\beta}= {\hat w}_{(\dot\alpha}^I {\hat\lambda}_{\dot\beta) I}\,, 
\\\nonumber 
N_I^{\,\, J} &=&  w^I_\alpha \lambda^\alpha_I +  w_{I \dot\alpha} \lambda^{I \dot\alpha}\,,
\\ \nonumber
\hat N_I^{\,\,J} &=&  {\hat w}_I^\alpha {\hat\lambda}_\alpha^I +  {\hat w}^{I \dot\alpha} {\hat\lambda}_{I \dot\alpha}\,.
\end{eqnarray}
They are gauge invariant under the transformations (\ref{gautra}). 
Finally, we can write the pure spinor ghost piece of the action
\begin{eqnarray}
 S_{\text{ghost}}&=&R^2\displaystyle\int d^2 z \Big(   {w}^{I}_\alpha \bar\nabla \lambda_{I}^\alpha+ {w}_{\dot\alpha I} \bar\nabla \lambda^{\dot\alpha I}+{\hat w}_{\alpha I} {\nabla} {\hat\lambda}^{ I \alpha}+{\hat w}_{\dot\alpha}^I {\nabla} {\hat\lambda}_{I}^{\dot\alpha} 
\\ \nonumber
&-&\eta^{(\alpha\beta)(\gamma\delta)}N_{\alpha\beta}{\hat N}_{\alpha\delta} - 
\eta^{(\dot\alpha\dot\beta)(\dot\gamma\dot\delta)}N_{\dot\alpha\dot\beta}{\hat N}_{\dot\gamma\dot\delta}-\eta^{I\,\,K}_{\,\,J\,\,L} N_I^{\,\, J}{\hat N}_K^{\,\, L}  \Big)\,,
\end{eqnarray}
where the bilinear metrics $\eta$ are given from (\ref{bil1}) and (\ref{bil2}) as
\begin{eqnarray}
 \eta^{(\alpha\beta)(\gamma\delta)}=\epsilon^{\alpha \gamma} \epsilon^{\beta \delta} + \epsilon^{\alpha \delta} \epsilon^{\beta\gamma},\;\;\;\;\;\eta^{I\,\,K}_{\,\,J\,\,L}=\delta^{~I}_{L}\delta_{J}^{~K} \,.
\end{eqnarray}

Putting everything together we get the pure spinor action for $AdS_4\times \mathbb{CP}^3$
\begin{eqnarray}\label{pure1}
S&=&R^2\displaystyle\int d^2z \Big[ \epsilon_{xy} J^x \bar{J}^y+\frac{1}{2} J_{IJ} \bar{J}^{IJ} + \frac{1}{4}\left( J_{\alpha I}{\bar J}^{\alpha I}+J_{\dot\alpha\bar I}{\bar J}^{\dot\alpha\bar I}\right)
+\frac{3}{4} \left( J_{\alpha \bar I}{\bar J}^{\alpha \bar I}+J_{\dot\alpha I}{\bar J}^{\dot\alpha I} \right)
\\ \nonumber
&&\;\;\;\;\;\;\;\;\;+  {w}^{I}_\alpha \bar\nabla \lambda_{I}^\alpha+ {w}_{\dot\alpha I} \bar\nabla \lambda^{\dot\alpha I}+{\hat w}_{\alpha I} {\nabla} {\hat\lambda}^{ I \alpha}+{\hat w}_{\dot\alpha}^I {\nabla} {\hat\lambda}_{I}^{\dot\alpha} 
\\ \nonumber
&&\;\;\;\;\;\;\;\;\;-\eta^{(\alpha\beta)(\gamma\delta)}N_{\alpha\beta}{\hat N}_{\gamma\delta}- 
\eta^{(\dot\alpha\dot\beta)(\dot\gamma\dot\delta)}N_{\dot\alpha\dot\beta}{\hat N}_{\dot\gamma\dot\delta}-\eta^{I\,\,K}_{\,\,J\,\,L} N_I^{\,\, J}{\hat N}_K^{\,\, L}  
\Big]\,,
\end{eqnarray}
The theory admits a BRST transformation with the following BRST charge
\begin{eqnarray}
 {Q}+\bar{{Q}}&=&\displaystyle\oint \left< dz \lambda J_3+d\bar z \hat\lambda {\bar J}_1 \right>
 \\ \nonumber
&=&\displaystyle\oint dz \left(
\lambda_{I \alpha} \hat J^{\alpha I}+ \lambda^{\dot\alpha I} \hat J_{\dot\alpha I} \right)
+
\displaystyle\oint d\bar z \left( {\hat\lambda}^{\alpha I} {\bar J}_{\alpha I}+ {\hat\lambda}^{\dot\alpha}_I {\bar J}_{\dot\alpha}^I \right)\,.
\end{eqnarray}

The general pure spinor action with ${\mathbb{Z}}_4$ discrete symmetry is invariant under the following BRST variations
\begin{eqnarray}
 \delta_B (J_0)&=&[J_3,\lambda]+[{\bar J}_1,\hat\lambda]
\\ \nonumber
 \delta_B (J_1)&=&\nabla \lambda +[J_2,\hat\lambda]
\\ \nonumber
 \delta_B (J_2)&=&[J_1,\lambda]+[J_3,\hat\lambda]
\\ \nonumber
 \delta_B (J_3)&=&\nabla\hat\lambda+[J_2,\lambda]
\\ \nonumber
 \delta_B (\lambda) &=& 0,\;\;\;\;\;\;\;\;\;\;\;\;  \delta_B (\hat\lambda) = 0\
\\ \nonumber
 \delta_B (\omega) &=& -J_3,\;\;\;\;\;\;\,\,  \delta_B (\hat\omega) = -{\bar J}_1\
\\ \nonumber 
 \delta_B (N) &=& [J_3,\lambda], \;\;\;\;  \delta_B (\hat N) = [{\bar J}_1,\hat\lambda]
\end{eqnarray}
where $\nabla Y=\partial Y+[J_0,Y]$ and $\bar\nabla Y=\bar\partial Y+[{\bar J}_0,Y]$. 
These can be written in the following form for the $AdS_4\times \mathbb{CP}^3$,
\begin{eqnarray}\label{BRST}
 \delta_B J_{\alpha\beta}&=& - 2 
 \lambda_{(\a I} \hat J^{I}_{\b)} - 2 J_{(\a I} \hat \lambda^{I}_{\b)} \,,
 \hspace{1cm}
 \delta_B J_{\dot\alpha\dot\beta}= - 2 
 \lambda_{(\dot \a}^I \hat J_{\dot \b)I } - 2 J_{(\dot\a}^I \hat \lambda_{\b)I} \,,
 \\ \nonumber
 \delta_B \hat J^{\alpha I}&=& 
(\nabla \hat\lambda)^{\alpha I} + J^{IJ} \lambda^\a_{J} + J^\a_{~\dot\a} \lambda^{\dot\a I}
,\hspace{.3cm} 
 \delta_B \hat J^{\dot\alpha}_I= 
 (\nabla \hat\lambda)^{\dot\alpha}_I + J_{IJ} \lambda^{\dot\alpha J} + 
 J_\a^{~\dot\a} \lambda^{\a}_I
\\ \nonumber 
 \delta_B J^{\alpha}_I&=& 
 (\nabla \lambda)^{\alpha}_I + J_{IJ} \hat\lambda^{\a J} + J^\a_{~\dot\a} \hat\lambda^{\dot\a}_I
,\hspace{.4cm} 
 \delta_B J^{\dot\alpha I}= 
  (\nabla \lambda)^{\dot \alpha I} + J^{IJ} \hat\lambda^{\dot\a}_J + J^\a_{~\dot\a} \hat\lambda^{\dot\a I}\,,
\\ \nonumber 
 \delta_B J_{\alpha\dot\beta}&=&  
 \lambda_{\a I} J^{I}_{\dot\b} + J_{\a I} \lambda^{I}_{\dot\b} + 
  \hat J_{\dot \beta I} \hat \lambda^{I}_{\a } + \hat \lambda_{\dot \beta I} \hat J^{I}_{\a } \,,
 \hspace{1cm}
\\\nonumber
 \delta_B J_{IJ}&=&  2\, \e^{\a\b} 
 \lambda_{\a [I} J_{J] \b} + 2\, \e^{\dot\a\dot\b} \hat J_{\dot\a [I} \hat\lambda_{J]\dot\b}
  \,,
 \\\nonumber
 \delta_B J^{IJ} &=&  2\, \e^{\a\b} 
 \lambda_{\a}^{[I} J^{J]}_{\b} + 2\, \e^{\dot\a\dot\b} \hat J_{\dot\a}^{[I} \hat\lambda^{J]}_{\dot\b}
  \,,
\\\nonumber
 \delta_B \omega_{\alpha}^I&=& - \hat J_\a^I\,,
\hspace{4cm} 
\delta_B \omega_{\dot\alpha I}= - \hat J_{\dot \a I}\,,
\\ \nonumber 
 \delta_B {\hat\omega}_{\alpha I}&=& - J_{\a I}\,,
 \hspace{3.8cm}
 \delta_B {\hat\omega}^{I}_{\dot\alpha}= - J_{\dot\a}^I\,,
\end{eqnarray}
the variations of $N_{\a\b}, N_{\dot\a\dot\b}, \hat N_{\a\b}, \hat N_{\dot\a\dot\b}$ can be easily derived by their definitions (\ref{lollo}). 
Using this notation, we can assign a further quantum number by assigning $0$ to $J_{\a\dot\a}$, $+1$ to $J^{IJ}$, $-1$ to $J_{IJ}$, 
$-1/2$ to $J_{\a I}, \hat J_{\dot \a, I}$ and $+1/2$ to $\hat J_{\a I}, J_{\dot \a, I}$. This is the center of $U(1)$ inside of $U(3)$. Notice that 
the symmetry is a $\mathbb Z_5$ symmetry. The action, the BRST transformations and the pure spinor conditions respect such a symmetry. 

\section{K\"ahler potential for the Grassmannian and the A-model}

Let's consider the Grassmannian coset $\frac{Ops(6|4)}{SO(6)\times Sp(4)}$ which is obtained out of the similar twisted coordinates $\Theta_A^x$ which was introduced by Berkovits for $AdS_5\times S^5$ \cite{Berkovits1}. A general K\"ahler potential on a coset $G/H$ was shown in \cite{itoh} to have the form 
\begin{equation}\label{kah}
 K(\Theta,\bar\Theta)=\frac{1}{2}\ln \det \left( \bar\xi(\bar\Theta) \xi(\Theta) \right)\,,
\end{equation}
 
where $\xi(\Theta)\in G/H$ is a representative of the coset $G/H$ where for any $h\in H$ and $g\in G$ satisfies
\begin{equation}
 g \xi(\Theta)=\xi(\Theta')h(\Theta,g)\,,
\end{equation}
 
Like in the case of the $G/H=\frac{PU(2,2|4)}{SU(4)\times SU(2,2)}$ coset \cite{Berkovits1}, for $G/H=\frac{Ops(6|4)}{SO(6)\times Sp(4)}$ also, there exists a gauging in which the coset representative can be written in the following form
\begin{equation}
 \xi=\mat{\ma_{4\times 4}&\Theta\\\bar\Theta&\ma_{6\times 6}},\;\;\;
\bar\xi=\mat{\ma_{4\times 4}&{\Theta}\\-\bar\Theta&\ma_{6\times 6}}\,,
\end{equation}

where here, $\Theta_A^x$ and $\bar\Theta_x^A$ are $4\times 6$ and $6\times 4$ fermionic matrices respectively.

Using the convention $i\bar\Theta=\Theta^\dagger$, the K\"ahler potential (\ref{kah}) can be written as

\begin{eqnarray}\label{kah1}
\nonumber
 K(\Theta,\bar\Theta)&=&\frac{1}{2}\ln\det \left[ \mat{\ma_{4\times 4}&\Theta\\\bar\Theta&\ma_{6\times 6}} 
\mat{\ma_{4\times 4}&{\Theta}\\-\bar\Theta&\ma_{6\times 6}}
 \right]
\\ \nonumber
&=&\frac{1}{2}\ln\det \left[ \mat{\ma_{4\times 4}-\Theta\bar\Theta& 0\\ 0 & \ma_{6\times 6}+\bar\Theta\Theta} \right]
\\ \nonumber
&=&\frac{1}{2}\ln\left[\det(\ma_{4\times 4}-\Theta\bar\Theta)\times\det(\ma_{6\times 6}+\bar\Theta\Theta) \right]
\\  
&=& \,\Tr \,\ln (\ma_{6\times 6}+\bar\Theta\Theta)
\end{eqnarray}

which in the last line we used the fact that
\begin{eqnarray}
 \Tr (\Theta\bar\Theta)^n=-\Tr (\bar\Theta\Theta)^n, \;\;\text{for }n>0\,,
\end{eqnarray}

One can easily show, for such a K\"ahler potential, exactly in the same way as it was shown in section (4.3) of \cite{Berkovits1}, that this N=2 action is conformal invariant, namely by computing the one-loop beta function
\begin{equation}
 R=\ln \det (\partial_\Theta\partial_{\bar\Theta} K)=0\,,
\end{equation}
which then the $N=2$ supersymmetry non-renormalization theorem implies its conformal invariance to all loops.

The worldsheet variables for this K\"ahler N=2 sigma-model on $\frac{Osp(6|4)}{SO(6)\times Sp(4)}$ are fermionic superfields $\Theta_A^x$ and $\bar\Theta_x^A$ where $A=1,\ldots ,6$ and $x=1,\ldots ,4$  
label fundamental representations of $SO(6)$ and $Sp(4)$ respectively. 
These $N=2$ chiral and anti-chiral superfields can be expanded in terms of the fields of the pure spinor superstring theory on the target $AdS_4\times \mathbb{CP}^3$ as follows
\begin{eqnarray}
\Theta_A^x(\kappa_{+},\kappa_{-})&=&\theta_A^x+\kappa_{+}Z_A^x+\kappa_{-}\bar{Y}_A^x+\kappa_{+}\kappa_{-}f_A^x\,,
\\ \nonumber
\bar\Theta^A_x(\bar\kappa_{+},\bar\kappa_{-})&=&\bar\theta^A_x+\bar\kappa_{+}\bar{Z}^A_x+\bar\kappa_{-}{Y}^A_x+
\bar\kappa_{+}\bar\kappa_{-}\bar{f}^A_x\,,
\end{eqnarray}
where $(\kappa_{+},\bar\kappa_{+})$ are left-moving and $(\kappa_{-},\bar\kappa_{-})$ are right-moving Grassmannian 
parameters of the worldsheet N=2 supersymmetry.

In this expansion, the 24 lowest components $\theta_A^x$ and $\bar\theta_x^A$ are 24 fermionic coordinates of the 
$\frac{Osp(6|4)}{U(3)\times SO(1,3)}$ supercoset which parametrizes the $AdS_4\times \mathbb{CP}^3$ superspace together with the 24 bosonic variables $Z_A^x$ and $\bar{Z}_x^A$ which are twistor-like variables combining the 10 spacetime 
coordinates of $AdS_4$ and $\mathbb{CP}^3$ with pure spinors $(\lambda_A^x,\bar\lambda^A_x)$ which their number was obtained in \cite{antonio} to be 14. 
They can be expressed explicitly as follows
\begin{eqnarray}
Z_A^x&=&H^x_{x'}(x_A)(\tilde{H}^{-1}({x_P}))_A^{A'}\,\lambda_{A'}^{x'}\,,
\\ \nonumber
\bar{Z}^A_x&=&(H^{-1}(x_A))_{x}^{x'}\tilde{H}_{A'}^A({x_P})\,\bar\lambda^{A'}_{x'}\,,
\end{eqnarray}

Here $H^x_{x'}(x_A)$ is a coset representative for the $AdS_4$ coset $\frac{Sp(4)}{SO(1,3)}$ 
 and $\tilde{H}_{A'}^A(x_P)$ 
is a coset representative for the $\mathbb{CP}^3$ coset $\frac{SO(6)}{U(3)}$.
Similarly, the conjugate twistor-like variables $Y_J^A$ and $\bar{Y}^J_A$ 
are constructed from the conjugate momenta to the pure spinors and
$f_A^x$ and $\bar{f}_x^A$ are auxiliary fields.

\section{From pure spinor to A-model}
Here we show that the same way Berkovits and Vafa \cite{BV} showed the equivalence of the A-model and the pure spinor superstrig for $AdS_5\times S^5$, 
we can show the existence of such an equivalence for any superscoset admitting a $\mathbb{Z}_4$ automorphism, as is the case also for the 
$AdS_4\times \mathbb{CP}^3$ supercoset.

\subsection{Pure spinor with $\mathbb{Z}_4$ automorphism and ``bonus`` symmetry}

Consider a supercoset $G/H$ which admits a $\mathbb{Z}_4$ automorphism under which its generators can be decomposed into invariant subspaces 
$\mathcal{H}_i,i=0\cdot 3$. The matter fields of the sigma model can be written in terms of the left-invariant currents 
$J=g^{-1}\partial g,\,\bar{J}=g^{-1}\bar\partial g$, where $g\in G$. The left-invariant currents are decomposed according to the invariant 
subspaces of the $\mathbb{Z}_4$ into $J=J_0+J_1+J_2+J_3$ as follows

\begin{equation}
\begin{matrix}
\mathcal{H}_0&\mathcal{H}_1&\mathcal{H}_2&\mathcal{H}_3\\J^{[AB]}&J^\alpha&J^M&J^{\hat\alpha}
\end{matrix}
\end{equation}

where the left-invariant current $J=g^{-1}\partial g$ is expanded by the generators of the superalgebra as
\begin{equation}
 J=\sum_{i=0}^3 J_i=J^{[AB]}T_{[AB]}+J^m T_m+J^\alpha T_\alpha+J^{\hat\alpha} T_{\hat\alpha}\,,
\end{equation}
here, $J^{[AB]}\in H$ are the spin connections of the supercoset and $J^m$ and ($J^\alpha,J^{\hat\alpha}$) are the bosonic and fermionic components of 
the supervielbein respectively. The generators of the supercoset are  $(T_{[AB]},T_m,T_\alpha,T_{\hat\alpha})$ which are the Lorentz generators, 
translations and fermionic generators respectively with the following non-zero structure constants
\begin{eqnarray}
 f_{mn}^{\;\;\;\; p},\;\;f_{mn}^{\;\;\;[AB]},\;\;f_{{[AB]}{[CD]}}^{\;\;\;\;\;\;\;\;\;[EF]},\;\;f_{\alpha\hat\beta}^{\;\;\;[AB]},\;\;f_{\alpha\beta}^{\;\;\; m}\,,
\end{eqnarray}

Besides the matter fields, the pure spinor action has  a ghost sector consisting of the pure spinors and their conjugate momenta
\begin{equation}
\lambda=\lambda^\alpha T_\alpha,\;\;\hat\lambda={\hat\lambda}^{\hat\alpha}T_{\hat\alpha},\;\;\omega=\eta^{\alpha\hat\alpha}
\omega_\alpha T_{\hat\alpha},\;\;\hat\omega=\eta^{\alpha\hat\alpha}{\hat\omega}_{\hat\alpha} T_{\alpha}\,,
\end{equation}
and the corresponding pure spinor currents $N=-\{\omega,\lambda\},\hat{N}=\{\hat\omega,\hat\lambda\}\in \mathcal{H}_0$ which generate the Lorentz 
transformations in the pure spinor variables.

The theory admits a BRST transformation with the following operator
\begin{equation}
 {Q}+\bar{{Q}}=\displaystyle\oint \left< dz \lambda J_3+d\bar z \hat\lambda {\bar J}_1 \right>\,,
\end{equation}
under which the fields transform as follows
\begin{eqnarray}
 \delta_B (J_0)&=&[J_3,\lambda]+[J_1,\hat\lambda]\,,
\\ \nonumber
 \delta_B (J_1)&=&\nabla \lambda +[J_2,\hat\lambda]\,,
\\ \nonumber
 \delta_B (J_2)&=&[J_1,\lambda]+[J_3,\hat\lambda]\,,
\\ \nonumber
 \delta_B (J_3)&=&\nabla\hat\lambda+[J_2,\lambda]\,,
\\ \nonumber
 \delta_B (\lambda) &=& 0,\;\;\;\;\;\;\;\;\;\;\;\;  \delta_B (\hat\lambda) = 0\,, \
\\ \nonumber
 \delta_B (\omega) &=& -J_3,\;\;\;\;\;\;\,\,  \delta_B (\hat\omega) = -{\bar J}_1 \,, \
\\ \nonumber 
 \delta_B (N) &=& [J_3,\lambda], \;\;\;\;  \delta_B (\hat N) = [{\bar J}_1,\hat\lambda]\,,
\end{eqnarray}
where $\nabla Y=\partial Y+[J_0,Y]$ and $\bar\nabla Y=\bar\partial Y+[{\bar J}_0,Y]$. These can also be written in the expanded form,
\begin{eqnarray}\label{BRST1}
\delta_B (J^{[AB]}) &=& J^{\hat\alpha} \lambda^\beta f_{\hat\alpha\beta}^{\;\;\;[AB]}+J^{\alpha}{\hat\lambda}^{\hat\beta} f_{\alpha\hat\beta}^{\;\;\;[AB]}\,,
\\ \nonumber
 \delta_B (J^m)&=&J^\alpha \lambda^\beta f_{\alpha\beta}^{\;\;\;m}+J^{\hat\alpha}{\hat\lambda}^{\hat\beta} f_{\hat\alpha\hat\beta}^{\;\;\;m}\,,
\\ \nonumber
 \delta_B (J^\alpha)&=& \nabla \lambda^\alpha +  J^m {\hat\lambda}^{\hat\alpha} f_{m \hat\alpha}^{\;\;\;\alpha}\,,
\\ \nonumber
 \delta_B (J^{\hat\alpha})&=& \nabla\hat\lambda^{\hat\alpha}+J^m\lambda^\alpha f_{m\alpha}^{\;\;\;\hat\alpha}\,.
\end{eqnarray}

The sigma model is invariant under the global transformations $\delta g=\Sigma g,\; \Sigma\in \cal G$ and under the BRST transformations, using the fact 
that $\left<AB\right>\neq 0$ only for $A\in \mathcal{H}_i$ and $B\in \mathcal{H}_{4-i}$. It can be written in the following form
\begin{equation}\label{z4}
 S=R^2\displaystyle\int d^2 z \left< \frac{1}{2} J_2 {\bar J}_2+\frac{1}{4} J_1{\bar J}_3+\frac{3}{4} J_3 {\bar J}_1 +w \bar\partial \lambda 
+ \hat w\partial\hat\lambda+N{\bar J}_0+{\hat N}J_0 - N\hat N\right>\,,
\end{equation}
for any supercoset admitting a ${\mathbb Z}_4$ automorphism including $AdS_5\times S^5$ and $AdS_4\times \mathbb{CP}^3$ examples 
(see also \cite{Adam:2006bt,Adam:2007ws} for non-critical examples based on different sets of pure spinor variables).

On top of the global bosonic isometry group $G_b$ of the supergroup $G$, the A-model action has a 'bonus' chiral symmetry exchanging left and right 
movers which appears in the sigma model as a symmetry between left and right moving  fermions $J^\alpha$ and $J^{\hat\alpha}$. Apparently (\ref{z4}) 
does not have such a symmetry because of the different coefficients of $J_1{\bar J}_3$ and $ J_3 {\bar J}_1$ terms. To promote the symmetry of (\ref{z4}), 
one can add an additional term to the action including a $-\frac{1}{2} J_3 {\bar J}_1$ to cancel the asymmetry of the fermionic currents together 
with its appropriate companion in order that the whole term stays a BRST-closed term,

\begin{eqnarray}\label{trivial}
 S_{trivial}&=&S_{m}+S_g \nonumber \\
 &=& \frac{R^2}{2}\displaystyle\int d^2z \left( C_{mn}J^m{\bar J}^n-<J_3 {\bar J}_1>+<\omega \bar\nabla \lambda + 
\hat\omega\nabla\hat\lambda - N\hat N>\right)
\\ \nonumber
&=&\frac{R^2}{2}\displaystyle\int d^2z \left( C_{mn}J^m{\bar J}^n+\eta_{\alpha\hat\beta} J^{\hat\beta} {\bar J}^\alpha+\omega_\alpha \bar\nabla \lambda^\alpha + 
\hat\omega_{\hat\alpha}\nabla\hat\lambda^{\hat\alpha} - \eta_{[AB][CD]} N^{[AB]}\hat N^{[CD]} \right)\,,
\end{eqnarray}

 where $S_g=\frac{R^2}{2} \int d^2z (\omega \bar\nabla \lambda + \hat\omega\nabla\hat\lambda - N\hat N)$ is exactly the ghost part of the original 
action (\ref{z4}) and $\eta_{XY}=<T_XT_Y>=Str(T_X T_Y)$. The requirement of BRST invariance of the $S_{trivial}$ will determine the unknown tensor $C_{mn}$.

Using the classical equations of motion
\begin{eqnarray}
 \nabla \hat\lambda-[N,\hat\lambda]=0,\;\;\;\;\bar\nabla\lambda-[\hat N,\lambda]=0\,,
\end{eqnarray}
and the identities $[ N,\lambda ] = [ \hat N,\hat\lambda ] =0 $ coming from the pure spinor constraints, it can be shown that under the BRST transformations (\ref{BRST1}), $S_{g}$ and $S_m$ vary as follows
\begin{eqnarray}
\label{1add} 
\delta_B(S_{g})&=&\frac{R^2}{2}\displaystyle\int d^2 z <-J_3\bar\partial\lambda-{\bar J}_1\partial\hat\lambda-J_3[{\bar J}_0,\lambda]-{\bar J}_1[J_0,\hat\lambda]>
\\ \nonumber
&=&\frac{R^2}{2}\displaystyle\int d^2 z \;\eta_{\alpha\hat\beta}(-J^{\hat\beta}\bar\nabla\lambda^\alpha+{\bar J}^\alpha\nabla {\hat\lambda}^{\hat\beta})\,,
\\ 
\label{2add} 
\delta_B(L_{m})&=&\frac{R^2}{2}\left[C_{mn}\left( J^\alpha \lambda^\beta f_{\alpha\beta}^{\;\;\;\;m}+J^{\hat\alpha}{\hat\lambda}^{\hat\beta}f_{\hat\alpha \hat\beta}^{\;\;\;\;m} \right){\bar J}^n+C_{mn} J^m\left( {\bar J}^\alpha \lambda^\beta f_{\alpha\beta}^{\;\;\;\;n} + {\bar J}^{\hat\alpha}{\hat\lambda}^{\hat\beta}f_{\hat\alpha\hat\beta}^{\;\;\;\;n}\right)\,\right.
\\ \nonumber 
&-&\left. \eta_{\alpha\hat\beta}\left( \nabla\hat\lambda^{\hat\beta}+J^m\lambda^\beta f_{m\beta}^{\;\;\;\;\hat\beta}  \right){\bar J}^\alpha+\eta_{\alpha\hat\beta} J^{\hat\beta}\left( \bar\nabla\lambda^{\alpha}+{\bar J}^n\hat\lambda^{\hat\alpha} f_{n\hat\alpha}^{\;\;\;\;\alpha} \right)\,\right],
\end{eqnarray}
which gives
\begin{eqnarray}\label{delsb}
\frac{1}{R^2} \delta_B(S_{trivial})&=&\frac{1}{2} C_{mn} J^m {\bar J}^\alpha\lambda^\beta f_{\alpha\beta}^{\;\;\;\; n}+\frac{1}{2}\eta_{\alpha\hat\beta} J^m{\bar J}^\alpha\lambda^\beta f_{m\beta}^{\;\;\;\;\hat\beta}\,,
\\ \nonumber 
&+&\frac{1}{2}C_{mn}{\bar J}^n J^{\hat\beta}{\hat\lambda}^{\hat\alpha}f_{\hat\beta\hat\alpha}^{\;\;\;\;m}+\frac{1}{2}\eta_{\alpha\hat\beta} {\bar J}^n J^{\hat\beta}{\hat\lambda}^{\hat\alpha}f_{n\hat\alpha}^{\;\;\;\;\alpha} 
\\ \nonumber
&+&\frac{1}{2}C_{mn}{\bar J}^n J^\alpha\lambda^\beta f_{\alpha\beta}^{\;\;\;\;m}+\frac{1}{2}C_{mn} J^m {\bar J}^{\hat\alpha}{\hat\lambda}^{\hat\beta} f_{\hat\alpha\hat\beta}^{\;\;\;\;n}
\\\nonumber
&=&0\,,
\end{eqnarray}	
which admits the following solution for $\delta_B(S_{trivial})=0$ after using the Jacobi identities for the structural constants
\begin{equation}\label{cmn}
 C_{mn}=\frac{1}{2}\frac{\eta_{\alpha\hat\beta}({\hat\lambda}^{\hat\alpha}f_{n\hat\alpha}^{\;\;\;\;\alpha})(\lambda^\beta f_{m\beta}^{\;\;\;\;\hat\beta})}{\eta_{\alpha\hat\beta}\lambda^\alpha{\hat\lambda}^{\hat\beta}}\,.
\end{equation}

The first and the second lines of (\ref{delsb})  vanish because of the identity $\eta_{\beta\hat\alpha}=Str\left(T_\beta T_{\hat\alpha}\right)=
f_{\alpha\beta}^{\,\,\,n} f_{n\hat\alpha}^{\,\,\alpha}$ 
and the terms in the last line vanish because of the following Jacobi identity,
\begin{equation}\label{jac1}
 f_{\alpha\gamma}^{\,\,m}f_{m\beta}^{\,\,\hat\beta}+f_{\alpha\beta}^{\,\,m}f_{m\gamma}^{\,\,\hat\beta}=f_{\beta\gamma}^{\,\,m}f_{m\alpha}^{\,\,\hat\beta}\,,
\end{equation}
which implies
\begin{equation}
 \lambda^\beta\lambda^\gamma\left(f_{\alpha\gamma}^{\,\,m}f_{m\beta}^{\,\,\hat\beta}+f_{\alpha\beta}^{\,\,m}f_{m\gamma}^{\,\,\hat\beta}\right)=0\,.
\label{jac2}
\end{equation}

So $S_{trivial}$ of (\ref{trivial}) with $C_{mn}$ given in (\ref{cmn}) is BRST-closed. We should also show that it is really a BRST-trivial 
term satisfying $S_{trivial}=Q\bar Q X$, up to the equations of motion. 
In order to do that, we introduce the antifields $w^*_\alpha$ and ${{\hat w}^*}_{\hat\alpha}$ which after adding the term
\begin{equation}
 R^2\displaystyle\int d^2 z \eta^{\alpha\hat\beta}{w^*}_\alpha{\hat w}^*_{\hat\beta}\,,
\end{equation}
the full action stay invariant under the new BRST transformations,
\begin{eqnarray}
 Q' w_\alpha&=&-\eta_{\alpha\hat\alpha}J^{\hat\alpha},\hspace{24.5mm} {\bar Q}' w_\alpha=w^*_\alpha\,,
\\\nonumber
Q'{\hat w}_{\hat\alpha}&=&{\hat w}^*_{\hat\alpha},\hspace{34mm} {\bar Q}'{\hat w}_{\hat\alpha}=-\eta_{\hat\alpha\alpha}{\bar J}^\alpha\,,
\\\nonumber
Q' w^*_\alpha&=&\eta_{\alpha\hat\alpha}(\nabla{\hat\lambda}^{\hat\alpha}-[N,\hat\lambda]^{\hat\alpha}),\hspace{4.5mm}\bar Q' w^*_\alpha=0\,,
\\\nonumber
Q' {\hat w}^*_{\hat\alpha}&=&0,\hspace{37mm} {\bar Q}' {\hat w}^*_{\hat\alpha}=\eta_{\hat\alpha\alpha}(\bar\nabla{\lambda}^{\alpha}-[\hat N,\lambda]^\alpha)\,,
\\\nonumber
Q' N&=&[J_3,\lambda],\hspace{30mm} {\bar Q}' N=[w^*,\lambda]\,,
\\\nonumber
Q' \hat N&=&[{\bar J}_1,\hat\lambda],\hspace{30mm} {\bar Q}' \hat N=[{\hat w}^*,\hat\lambda]\,,
\end{eqnarray}
which this BRST transformation is nilpotent off-shell instead of being nilpotent up to the equations of motion. Now consider the following identities
\begin{eqnarray}
 \label{qq1}
Q'{\bar Q}' \left(C_{mn}J^m{\bar J}^n\right)&=&C_{mn} \left\lbrace Q'{\bar Q}' (J^m){\bar J}^n +Q'(J^m){\bar Q}'({\bar J}^n) +{\bar Q}'(J^m) Q'({\bar J}^n) + J^mQ'{\bar Q}'({\bar J}^n) \right\rbrace
\nonumber \\ \nonumber
&=&C_{mn} \left\lbrace \nabla {\hat\lambda}^{\hat\alpha}\,{\hat\lambda}^{\hat\beta} f_{\hat\alpha\hat\beta}^{\;\;\;\; m}{\bar J}^n+
J^m   \bar\nabla {\lambda}^{\alpha}\,{\lambda}^{\beta} f_{\alpha\beta}^{\;\;\;\; n} \right\rbrace
\\ \nonumber
&+& C_{mn} \left\lbrace J^p{\bar J}^n\lambda^\alpha{\hat\lambda}^{\hat\beta}f_{p\alpha}^{\;\;\;\;\hat\alpha}f_{\hat\alpha\hat\beta}^{\;\;\;\;m}+J^m{\bar J}^p \lambda^\alpha {\hat\lambda}^{\hat\beta}f_{p\alpha}^{\;\;\;\;\hat\alpha}f_{\hat\alpha\hat\beta}^{\;\;\;\;n} \right\rbrace
\\ \nonumber 
&+& C_{mn} \left\lbrace J^\alpha{\bar J}^{\hat\alpha} \lambda^\beta{\hat\lambda}^{\hat\beta} f_{\alpha\beta}^{\;\;\;\;m}f_{\hat\alpha\hat\beta}^{\;\;\;\;n} +J^{\hat\alpha}{\bar J}^{\alpha} {\hat\lambda}^{\hat\beta}{\lambda}^{\beta} f_{\hat\alpha\hat\beta}^{\;\;\;\;m} f_{\alpha\beta}^{\;\;\;\;n} \right\rbrace
\\
&=&2C_{mn}J^m{\bar J}^n \left(\eta \lambda\hat\lambda\right)\, ,
\end{eqnarray}
\begin{eqnarray}
\label{qq2}
Q'{\bar Q}' \left(N\hat N\right) &=&Q'{\bar Q}'(N) \hat N+Q'(N){\bar Q}'(\hat N)+{\bar Q}'(N)Q' (\hat N)+NQ'{\bar Q}'(\hat N)
\\ \nonumber
&=&[(\nabla\hat\lambda-[N,\hat\lambda]),\lambda]\hat N+[J_3,\lambda][{\bar J}_1,\hat\lambda]+[w^*,\lambda][{\hat w}^*,\hat\lambda]+N[(\bar\nabla\lambda-[\hat N,\lambda]),\hat\lambda]\,,
\end{eqnarray}
and,
\begin{eqnarray}
\label{qq3}\nonumber
Q'{\bar Q}' \left((\omega\lambda)(\hat\omega\hat\lambda)\right)&=&Q'{\bar Q}'(\omega\lambda)(\hat\omega\hat\lambda)+ Q'(\omega\lambda){\bar Q}'(\hat\omega\hat\lambda)+{\bar Q}'(\omega\lambda){ Q}'(\hat\omega\hat\lambda)+(\omega\lambda){ Q}'{\bar Q}'(\hat\omega\hat\lambda)
\\ \nonumber
&=&\frac{1}{2}[(\nabla\hat\lambda-[N,\hat\lambda]),\lambda](\hat w\hat\lambda)+\frac{1}{4}[J_3,\lambda][{\bar J}_1,\hat\lambda]+\frac{1}{4}[w^*,\lambda][{\hat w}^*,\hat\lambda] \nonumber\\
&+&\frac{1}{2}(w\lambda)[(\bar\nabla\lambda-[\hat N,\lambda]),\hat\lambda]\,,
\end{eqnarray}
to get these identities, we used the equation of motions, (\ref{delsb}), (\ref{jac1}) 
and (\ref{jac2}) together with the following Jacobi identity,
\begin{eqnarray}\label{JI}
f_{M\underline\alpha}^{\;\;\;\;\underline\beta}f_{N\underline\beta}^{\;\;\;\;\underline\gamma}-f_{N\underline\alpha}^{\;\;\;\;\underline\beta}f_{M\underline\beta}^{\;\;\;\;\underline\gamma}=f_{MN}^{\;\;\;\;P}f_{P \underline{\alpha}}^{\;\;\;\;\underline{\beta}}\,,
\end{eqnarray}
where $M,N,\cdots=\lbrace m,[mn]\rbrace$ and $\underline\alpha,\underline\beta,\cdots=\lbrace \alpha,\hat\alpha\rbrace$. 
From (\ref{qq1}), (\ref{qq2}) and (\ref{qq3}) one can see that there exists a linear combination of them such that $S_{trivial}=Q\bar Q X$ up to the anti-ghost term, 
that is up to the momenta equations of motion,
\begin{equation}
 X=\frac{1}{2}\displaystyle\int d^2 z \frac{1}{\eta_{\alpha\hat\alpha}\lambda^\alpha{\hat\lambda}^{\hat\alpha}}\left[\frac{1}{4}C_{mn}J^m{\bar J}^n+\frac{1}{4}(\omega\lambda)(\hat\omega\hat\lambda)-\frac{1}{8}N\hat N \right]\,.
\end{equation}
The sigma model action after adding $S_{trivial}$ becomes
\begin{eqnarray}\label{ps1}
 S_b&=&\frac{R^2}{2}\displaystyle\int d^2 z \Big[\left(\frac{1}{2}\frac{\eta_{\alpha\hat\beta}({\hat\lambda}^{\hat\alpha}f_{n\hat\alpha}^{\;\;\;\;\alpha})(\lambda^\beta f_{m\beta}^{\;\;\;\;\hat\beta})}{\eta_{\alpha\hat\beta}\lambda^\alpha{\hat\lambda}^{\hat\beta}}+\eta_{mn}\right)
j^m{\bar J}^n
\\\nonumber
&&\;\;\;\;\;\;\;\;\;\;\;\;\;\;+\frac{1}{2} < J_3 {\bar J}_1-J_1 {\bar J}_3 +\omega\bar\nabla\lambda+\hat\omega\nabla\hat\lambda-N\hat N > \Big]\,
\end{eqnarray}
The analysis follows the considerations in the literature, but it is derived in a very general way. 

\subsection{Mapping pure spinor to A-model}

In order to relate $S_b$ and the A-model action, we should write the supercoset element $g(x,\theta,\bar\theta)\in\frac{G}{H}$ in terms of 
the Grassmannian coset element $G(\theta,\bar\theta)\in \frac{G}{G_b}$. 

We can define the following bosonic twisted variables out of the bosonic coset elements $H(x)\in\frac{G_b}{H}$ and the pure spinors in this way

\begin{eqnarray}\label{twisted}
 Z^\alpha&=&[H,\lambda]=H^{[AB]}(x)\lambda^\beta f_{[AB]\beta}^{\;\;\;\;\;\;\;\;\alpha}
\\ \nonumber
 {\bar Z}^{\hat\alpha}&=&[H^{-1},\hat\lambda]=(H^{-1})^{[AB]}(x){\hat\lambda}^{\hat\beta} f_{[AB]\hat\beta}^{\;\;\;\;\;\;\;\;\hat\alpha}
\\ \nonumber
 { Y}^{\hat\alpha}&=&[H^{-1}, w]=(H^{-1})^{[AB]}(x)\eta^{\beta\hat\beta}{w}_{\beta} f_{[AB]\hat\beta}^{\;\;\;\;\;\;\;\;\hat\alpha}
\\ \nonumber
 {\bar Y}^\alpha&=&[H,\hat w]=H^{[AB]}(x)  \eta^{\beta\hat\beta}{\hat w}_{\hat\beta} f_{[AB]\beta}^{\;\;\;\;\;\;\;\;\alpha}
\end{eqnarray}

Supercoset element $g$ can be parametrized as follows

\begin{equation}\label{gG}
 g(x,\theta,\bar\theta)=G(\theta,\bar\theta)H(x)
\end{equation}

where $G(\theta,\bar\theta)=e^{\theta^\alpha T_\alpha+{\bar\theta}^{\hat\alpha} T_{\hat\alpha}}$ and $H(x)=e^{x^m T_m}$ in which 
$(T_m,T_\alpha,T_{\hat\alpha})$ are the generators of the supercoset $G/H$.

According to (\ref{gG}), we can decompose the left-invariant currents $J=g^{-1} \partial g$. The pure spinor action can be written 
into $H$ and $G$ components, corresponding to the purely bosonic part and purely fermionic part of the supercoset as follows
\begin{eqnarray}
 J=H^{-1}\partial H+H^{-1}(G^{-1}\partial G)H
\end{eqnarray}

Its componets $J=J^m T_m+J^{[AB]}T_{[AB]}+J^\alpha T_\alpha+J^{\hat\alpha}T_{\hat\alpha}$ can be written as
\begin{eqnarray}
 J^M&=&(H^{-1}\partial H)^M+(H^{-1})^M(G^{-1}\partial G)^P H^{Q}f_{NP}^{\;\;\;\;R}f_{RQ}^{\;\;\;\;M}
\\
J^{\underline\alpha}&=&(H^{-1})^M(G^{-1}\partial G)^{\underline\beta} H^N f_{M\underline\beta}^{\;\;\;\;\underline\gamma}f_{\underline\gamma N}^{\;\;\;\;\underline\alpha}
\end{eqnarray}
where $M,N,\cdots=\{m,[AB]\}$ and $\underline{\alpha},\underline{\beta},\cdots=\{\alpha,\hat\alpha\}$.

The A-model action can be written in terms of the fermionic superfields $(\Theta^\alpha,{\bar\Theta}^{\hat\alpha})$ which was defined before as 
$S=\int \Tr \ln[1+\bar\Theta\Theta]$. Here we assume that for the Grassmannian supercoset $G/G_b$, there exist a gauging in which the supercoset 
elements $G$ can be written in the following form
\begin{equation}
 G^m=\ma,\;\;\;G^{[AB]}=\ma,\;\;\;G^{\alpha}=\theta^\alpha,\;\;\;G^{\hat\alpha}={\bar\theta}^{\hat\alpha}
\end{equation}
Finally, the A-model action, after integration over the auxiliary fields can be written in this form
\begin{eqnarray}\label{amodel}
 S_A&=&t \displaystyle\int d^2 z\Big[(G^{-1}\partial G)(G^{-1}\bar\partial G)+Y\bar\nabla Z+\bar Y\nabla\bar Z-(YZ)(\bar Z\bar Y)\Big]
\\\nonumber
&=&t\displaystyle\int d^2 z [\eta_{\alpha\hat\alpha}(G^{-1}\partial G)^\alpha(G^{-1}\bar\partial G)^{\hat\alpha}+\eta_{MN}(G^{-1}\partial G)^M(G^{-1}\bar\partial G)^{N}
\\ \nonumber
&+& \eta_{\alpha\hat\alpha}Y^{\hat\alpha}(\bar\nabla Z)^\alpha+\eta_{\alpha\hat\alpha}{\bar Y}^\alpha(\nabla\bar Z)^{\hat\alpha}-\eta_{mn}f_{\alpha\hat\alpha}^{\;\;\;\;m}f_{\beta\hat\beta}^{\;\;\;\;n}\left[(Y^{\hat\alpha}Z^{\alpha})({\bar Z}^{\hat\beta}{\bar Y}^{\beta})+(Z^\alpha Y^{\hat\alpha})({\bar Y}^{\beta}{\bar Z}^{\hat\beta})\right]]
\end{eqnarray}
where,
\begin{eqnarray}\label{cdev}
 (\bar\nabla Z)^\alpha&=& \bar\partial Z+[G^{-1}\bar\partial G,Z]
\\\nonumber
&=&\bar\partial Z^\alpha+(G^{-1}\bar\partial G)^{[AB]} Z^\beta f_{[AB]\beta}^{\;\;\;\;\;\;\alpha}
\\ \nonumber
 (\nabla \bar Z)^{\hat\alpha}&=& \partial \bar Z+[G^{-1}\partial G,\bar Z]
\\\nonumber
&=&\partial {\bar Z}^{\hat\alpha}+(G^{-1}\partial G)^{[AB]} Z^{\hat\beta} f_{[AB]\hat\beta}^{\;\;\;\;\;\;\hat\alpha}
\end{eqnarray}

To relate the pure spinor action (\ref{ps1}) and the A-model action (\ref{amodel}), we use the explicit form of the twisted variables (\ref{twisted}). 
Using (\ref{twisted}) and Jacobi identity (\ref{JI}), one can write
\begin{eqnarray}
Y\bar\partial Z&=&[H^{-1},w]\bar\partial\left([H,\lambda]\right)
\\\nonumber
&=&[H^{-1},w]\left([\bar\partial H,\lambda]+[H,\bar\partial\lambda]\right)
\\\nonumber
&=&w\bar\partial\lambda+[H^{-1}\bar\partial H,w\lambda]
\\\nonumber
&=&w\bar\partial\lambda+[H^{-1}\bar\partial H,w\lambda]+[H^{-1}(G^{-1}\bar\partial G)H,w \lambda]-[H^{-1}(G^{-1}\bar\partial G)H,w \lambda]
\\\nonumber
&=&w\bar\partial\lambda+[\bar J,w\lambda]-[(G^{-1}\bar\partial G),YZ]
\end{eqnarray}

which after using (\ref{cdev}), we get

\begin{eqnarray}
Y\bar\nabla Z&=&w\bar\partial\lambda+[\bar J,w\lambda]
\\\nonumber
&=&w_\alpha\bar\partial\lambda^\alpha+{\bar J}^{[AB]}w_\alpha\lambda^\beta f_{[AB] \beta}^{\,\,\,\,\,\,\,\,\,\,\,\alpha}+\eta_{mn}\eta^{\alpha\beta} {\bar J}^m w_\alpha \lambda^{\gamma} f_{\gamma\beta}^{\,\,\,\,n}
\\\nonumber
&=&w_\alpha\bar\nabla\lambda^\alpha+\eta_{mn}\eta^{\alpha\beta} {\bar J}^m w_\alpha \lambda^{\gamma} f_{\gamma\beta}^{\,\,\,\,n}
\end{eqnarray}

similarly, one can see that

\begin{eqnarray}
\bar Y\nabla \bar Z&=&\hat w\partial\hat\lambda+[ J,\hat w\hat\lambda]
\\\nonumber
&=&{\hat w}_{\hat\alpha}\partial{\hat\lambda}^{\hat\alpha}+{J}^{[AB]}{\hat w}_{\hat\alpha}{\hat\lambda}^{\hat\beta} f_{[AB] \hat\beta}^{\,\,\,\,\,\,\,\,\,\,\,\hat\alpha}+\eta_{mn}\eta^{\hat\alpha\hat\beta} { J}^m {\hat w}_{\hat\alpha} {\hat\lambda}^{\hat\gamma} f_{\hat\gamma\hat\beta}^{\,\,\,\,n}
\\\nonumber
&=&{\hat w}_{\hat\alpha}\nabla{\hat\lambda}^{\hat\alpha}+\eta_{mn}\eta^{\hat\alpha\hat\beta} { J}^m {\hat w}_{\hat\alpha} {\hat\lambda}^{\hat\gamma} f_{\hat\gamma\hat\beta}^{\,\,\,\,n}
\end{eqnarray}
 the last term simplifies as follows
\begin{eqnarray}
(YZ)(\bar Z\bar Y)&=&\left([H^{-1},w][H,\lambda]\right)\left([H^{-1},\hat\lambda][H,\hat w]\right)
\\\nonumber
&=&(w\lambda)(\hat w\hat\lambda)
\\\nonumber
&=&\eta^{[AB][CD]}\left(f_{\alpha [AB]}^{\;\;\;\;\;\;\;\beta}w_\beta\lambda^\alpha\right)\left({\hat w}_{\hat\beta}{\hat\lambda}^{\hat\alpha}f_{\hat\alpha [CD]}^{\;\;\;\;\;\;\;\hat\beta}\right)
-\eta_{mn}\left(\eta^{\alpha\gamma}f_{\alpha\beta}^{\,\,\,\,m}w_\gamma \lambda^\beta \right)\left(\eta^{\hat\alpha\hat\gamma}f_{\hat\alpha\hat\beta}^{\,\,\,\,n}{\hat w}_{\hat\gamma}{\hat\lambda}^{\hat\beta}\right)
\end{eqnarray}

Putting everything together, we obtain the A-model action in terms of the pure spinor fields
\begin{eqnarray}\label{eqs1}
 S_A&=&t\displaystyle\int d^2 z \Big[\frac{1}{2}\eta_{\alpha\hat\beta}(J^{\hat\beta}{\bar J}^\alpha-J^\alpha{\bar J}^{\hat\beta}) +w \bar\nabla\lambda+{\hat w}\nabla{\hat\lambda}-N\hat N
\\\nonumber
&&\;\;\;\;\;\;\;\;\;\;\;\;\;\;+\eta^{\alpha\beta} {\bar J}^m w_\alpha \lambda^{\gamma} f_{m\gamma}^{\,\,\,\alpha}+\eta^{\hat\alpha\hat\beta} { J}^m {\hat w}_{\hat\alpha} {\hat\lambda}^{\hat\gamma} f_{m\hat\gamma}^{\,\,\,\hat\alpha}-\eta_{mn}\left(\eta^{\alpha\gamma}f_{\alpha\beta}^{\,\,\,\,m}w_\gamma \lambda^\beta \right)\left(\eta^{\hat\alpha\hat\gamma}f_{\hat\alpha\hat\beta}^{\,\,\,\,n}{\hat w}_{\hat\gamma}{\hat\lambda}^{\hat\beta}\right)\Big]
\end{eqnarray}

The equations of motion for $w$ and $\hat w$ comes from the variation of the action under the transformations $\delta w_\alpha=f_{\alpha\beta}^{\,\,\,\,m}  \lambda^\beta \Lambda_m$ and $\delta{\hat w}_{\hat\alpha}=f_{\hat\alpha\hat\beta}^{\,\,\,\,m}{\hat\lambda}^{\hat\beta}{\tilde\Lambda}_m$, as follows
\begin{eqnarray}\label{eqw}
 (f_{m\alpha}^{\,\,\,\,\hat\delta}\lambda^\alpha)\left({\bar J}^m-\eta^{\hat\beta\hat\gamma}f_{\hat\beta\hat\alpha}^{\,\,\,\,m}{\hat w}_{\hat\gamma}{\hat\lambda}^{\hat\alpha}\right)&=&0
\\\nonumber
 (f_{m\hat\alpha}^{\,\,\,\,\delta}{\hat\lambda}^{\hat\alpha})\left({ J}^m-\eta^{\beta\gamma}f_{\beta\alpha}^{\,\,\,\,m}{ w}_{\gamma}{\lambda}^{\alpha}\right)&=&0
\end{eqnarray}

After inserting these equations of motion into (\ref{eqs1}), the second line of (\ref{eqs1}) produces the kinetic term for the bosonic Maurer-Cartan currents,
\begin{equation}
t\displaystyle\int d^2 z \Big[\frac{1}{2}\frac{\eta_{\alpha\hat\beta}({\hat\lambda}^{\hat\alpha}f_{n\hat\alpha}^{\;\;\;\;\alpha})(\lambda^\beta f_{m\beta}^{\;\;\;\;\hat\beta})}{\eta_{\alpha\hat\beta}\lambda^\alpha{\hat\lambda}^{\hat\beta}}+\eta_{mn}\Big]  J^n{\bar J}^m 
\end{equation}

Then the action (\ref{eqs1}), becomes
\begin{eqnarray}
 S&=&t\displaystyle\int d^2 z \Big[\left(\frac{1}{2}\frac{\eta_{\alpha\hat\beta}({\hat\lambda}^{\hat\alpha}f_{n\hat\alpha}^{\;\;\;\;\alpha})(\lambda^\beta f_{m\beta}^{\;\;\;\;\hat\beta})}{\eta_{\alpha\hat\beta}\lambda^\alpha{\hat\lambda}^{\hat\beta}}+\eta_{mn}\right)  J^n{\bar J}^m +\frac{1}{2}\eta_{\alpha\hat\beta}(J^{\hat\beta}{\bar J}^\alpha-J^\alpha{\bar J}^{\hat\beta} )
\\ \nonumber
&&\;\;\;\;\;\;\;\;+w \bar\nabla\lambda+{\hat w}\nabla{\hat\lambda}-N\hat N \Big]
\end{eqnarray}
which coincides with the action (\ref{ps1}) after identifying $t=\frac{1}{2}R^2$.

\section{Linear gauged $\sigma-$model for $AdS_4\times \mathbb{CP}^3$}

Similarly to the non-linear sigma model of the $AdS_5\times S^5$ which was studied by Berkovits and Vafa in \cite{BV}, we can write a linear 
gauged sigma model for the non-linear sigma model for $AdS_4\times \mathbb{CP}^3$ which was given in the previous section. 

The 2-dimensional N=(2,2) linear gauged sigma model can be described by a set of matter fields which are chiral and antichiral superfields 
$\Phi_R^\Sigma$ and ${\bar\Phi}_\Sigma^R$ gauged under the real worldsheet superfield $V_S^R$ taking value in the $SO(6)$ gauge group where $R,S,...=1,\ldots ,6$ are gauge field indices and $\Sigma=(x,A)$ is a global $Osp(6|4)$ index. We can take $\Phi_R^x$ to be fermionic 
while $\Phi_R^A$ are bosonic superfields.

The gauged linear sigma model action can be written in a $Osp(6|4)$ invariant way as
\begin{equation}
 S=\displaystyle\int d^2z \displaystyle\int d^4\kappa \left[ {\bar\Phi}_\Sigma^S (e^V)_S^R {\Phi}^\Sigma_R+t \,\Tr V+\frac{1}{e^2}\Sigma^2 \right]
\end{equation}
where $\Sigma=\bar{D} D V$ is the field strength of the gauge field $V$ and is a twisted chiral superfield.

As it is clear from the matter content of the theory, it contains 24 fermions and 36 bosons and so the theory actually has conformal anomaly 
if we ask the bosons and fermions to be gauged in the same representation of the gauge group as we did. But still the theory has 
a conformal IR fixed point corresponding to the large volume and gauge coupling limit which after integrating out the auxiliary 
equations of motion for the gauge field we obtain the non-linear sigma model (when $e\rightarrow \infty$)

\begin{equation}
 S=t\displaystyle\int d^2 z \displaystyle\int d^4\kappa  \Tr \left[ {\bar\Phi}_\Sigma^R \Phi_S^\Sigma \right]
\end{equation}

which can be rewritten in terms of the meson fields $\Theta_A^x$ and $\bar\Theta_x^A$ defined as
\begin{eqnarray}
 \Theta_A^x\equiv \Phi_R^x (\Phi^{-1})^R_A,\;\;\;\;\;\;\bar\Theta^A_x\equiv ({\bar\Phi}^{-1})_R^A\bar\Phi^R_x 
\end{eqnarray}
which gives exactly the A-model sigma model which was obtained from the pure spinor string for $AdS_4\times \mathbb{CP}^3$ as
\begin{equation}
 S=t \displaystyle\int d^2z \displaystyle\int d^4\kappa \Tr\,\ln\left[1+\bar\Theta\Theta\right]  
\end{equation}

The FI parameter corresponds to the K\"ahler parameter of the supercoset 
Grassmannian target space $\frac{Osp(6|4)}{SO(6)\times Sp(4)}$. 

\subsection{Vacua of the gauged linear sigma model and zero radius limit}

The small radius limit of the gauged linear sigma-model is convenient to study the perturbative regime of the gauge theory since the introduction 
of the Coulomb branch, because of the presence of the gauge group which is an additional degree of freedom in the gauged linear sigma model with 
respect to non-linear sigma model, resolves the singularity of the non-linear sigma-model in the small radius limit. To study different phases of 
the theory, we should solve the D-term equations comming from the gauged linear sigma-model. It is enough to focus on the fields which have 
conformal weight zero because they are the only fields which can get non-zero expectation value. 
We analyze the gauged linear $\sigma$-model following the standard techniques of \cite{Witten:1993yc} and \cite{Seki:2006cj}.

 The gauge superfield $V_S^R$ in Wess-Zumino gauge can be expanded as

\begin{equation}
 V_S^R=\sigma_S^R \kappa_+{\bar\kappa}_-+{\bar\sigma}_S^R {\kappa}_+{\bar\kappa}_++\ldots+\kappa_+\kappa_-{\bar\kappa}_+{\bar\kappa}_-D_S^R
\end{equation}
 
similarly we can expand the fermionic and bosonic superfields as follows

\begin{eqnarray}
 \Phi_R^\Sigma=\phi_R^\Sigma+\kappa_+ \psi_R^\Sigma+\ldots,\;\;\;\;\;
 {\bar\Phi}^R_\Sigma={\bar\phi}^R_\Sigma+{\bar\kappa}_- {\bar\psi}^R_\Sigma+\ldots
\end{eqnarray}

where we just keep the components which will have zero conformal weight after the A-twist because they are the only fields which can attain nonzero 
expectation value and so can appear in the D-term equations. Here the index $\Sigma$ refers to both $x$ and $A$ indices. 
Note that $(\phi_R^A,\psi_R^x,{\bar\phi}_A^R,{\bar\psi}_x^R)$ are bosonic and  $(\phi_R^x,\psi_R^A,{\bar\phi}_x^R,{\bar\psi}_A^R)$ are fermionic fields. 

Using the vector superfield and the usual superderivatives $D_{\pm}$ and ${\bar D}_{\pm}$, one can define the covariant superderivatives as follows
\begin{eqnarray}
 {\mathcal{D}}_{\pm}=e^{-V}D_{\pm}e^{+V},\;\;\;{\bar{\mathcal{D}}}_{\pm}=e^{+V}{\bar D}_{\pm}e^{-V}
\end{eqnarray}

Then the field strength $\Sigma$ which is a twisted chiral superfield is constructed as follows

\begin{eqnarray}
 \Sigma&=&\{{\bar{\mathcal{D}}}_+,{\mathcal{D}}_-\}
\\ \nonumber
&=& \sigma+\ldots+\kappa_+\kappa_-{\bar\kappa}_+{\bar\kappa}_- (D^m D_m \sigma+[\sigma,[\sigma,\bar\sigma]]+i[\partial^m v_m,\sigma])
\end{eqnarray}

which produces the following gauge field kinetic term in the Lagrangian

\begin{eqnarray}
 L_{gauge}&=&-\frac{1}{e^2}\displaystyle\int d^4\kappa \Tr \bar\Sigma\Sigma
\\ \nonumber
&=& \frac{1}{e^2} \Tr\left(-D_i \bar\sigma D^i \sigma -\frac{1}{2} [\sigma,\bar\sigma]^2 +\ldots \right)
\end{eqnarray}

and also we have the FI term $L_{D,\theta}$,

\begin{eqnarray}
 L_{D,\theta}&=& {it} \left.\displaystyle\int d\kappa_+ d{\bar\kappa}_- \Tr \Sigma\right|_{\kappa_-={\bar\kappa}_+=0}-{i\bar t} \left.\displaystyle\int d\kappa_- d{\bar\kappa}_+ \Tr \bar\Sigma\right|_{\kappa_+={\bar\kappa}_-=0}
\\ \nonumber
&=&\Tr \left( -rD +\frac{\theta}{2\pi} v_{01} \right)
\end{eqnarray}

Now we can consider the matter part of the gauged linear sigma model consisting of the kinetic terms for the fermionic and bosonic superfields 
which carries the kinetic and interaction terms for the bosonic and fermionic fields,

\begin{eqnarray}
 L_{kin}^b&=& \displaystyle\int d^4 \kappa  {\bar\Phi}_A^R e^V \Phi_R^A
\\ \nonumber
&=& -({\bar D}_j {\bar\phi}_A^R)(D^j \phi_R^A)+{\bar F}^R_A F_R^A-{\bar\phi}_A^S \{\sigma,\bar\sigma \}^R_S \phi_R^A+{\bar\phi}_A^S D_S^R \phi^A_R + \ldots 
\end{eqnarray}

Similarly we can write the kinetic term for the fermionic chiral superfields,

\begin{eqnarray}
 L_{kin}^f&=& \displaystyle\int d^4 \kappa  {\bar\Phi}_x^R e^V \Phi_R^x
\\ \nonumber
&=& -({\bar D}_j {\bar\phi}_x^R)(D^j \phi_R^x)+{\bar F}^R_x F_R^x-{\bar\phi}_x^S \{\sigma,\bar\sigma \}^R_S \phi_R^x+{\bar\phi}_x^S D_S^R \phi^x_R + \ldots 
\end{eqnarray}

We can see that $\{\sigma,\bar\sigma\}$ appears as the mass for the matter fields and so whenever $\sigma$ gets VEV, the matter fields 
become massive and can be integrated out in the effective theory as is happening in the Coulomb phase. 

The potential of the theory can be written as,
\begin{eqnarray}
 L_V&=&\frac{1}{2e^2} \Tr D^2-r\Tr D+{\bar\phi}_x^S D_S^R \phi^x_R +{\bar\phi}_A^S D_S^R \phi^A_R
\\ \nonumber
&-&\frac{1}{2e^2} \Tr [\sigma,\bar\sigma]^2-{\bar\phi}_x^S \{\sigma,\bar\sigma \}^R_S \phi_R^x-{\bar\phi}_A^S \{\sigma,\bar\sigma \}^R_S \phi_R^A
\end{eqnarray}
which after eliminating the D-field by using the following D-term equation
\begin{equation}
 D_R^S={\bar\phi}_x^S  \phi^x_R +{\bar\phi}_A^S  \phi^A_R - r \delta^S_R
\end{equation}
one obtains the potential 
\begin{eqnarray}
 V&=&\frac{e^2}{2} \left[{\bar\phi}_x^S  \phi^x_R +{\bar\phi}_A^S  \phi^A_R - r \delta^S_R\right]\left[{\bar\phi}_x^R  \phi^x_S +{\bar\phi}_A^R  \phi^A_S - r 
\delta^R_S \right]
\\ \nonumber
&+& \frac{1}{2e^2} \Tr [\sigma,\bar\sigma]^2+{\bar\phi}_x^S \{\sigma,\bar\sigma \}^R_S \phi_R^x+{\bar\phi}_A^S \{\sigma,\bar\sigma \}^R_S \phi_R^A
\end{eqnarray}
The space of the classical vacua is given by putting the potential to zero up to gauge transformations. We can study the vacua in two regimes, 
when $r>0$ and not small, the constraint $V=0$ implies that $\sigma=0$ which implies the following condition as the classical vacua for the matter fields
\begin{equation}
 D_R^S={\bar\phi}_x^S  \phi^x_R +{\bar\phi}_A^S  \phi^A_R - r \delta^S_R=0
\end{equation}
It means actually that the vectors $(\phi^x_R,\psi_R^A)$ for any $R=1,...,4$ are orthonormal.  Any such vector, after diagonalization, 
is subject to the constraint 
\begin{equation}
\sum_{A=1}^6{\bar\phi}_A \phi^A + \sum_{x=1}^4{\bar\phi}_x  \phi^x  =r
\end{equation}
which defines a supersphere ${\mathbb S}^{(5|4)}$.\footnote{The conditions for a supermanifold of being a super-Ricci flat are discussed in \cite{Grassi:2006cd}.}  
The space of classical vacua is the gauge invariant subspace of the product of such vectors 
\cite{cv}
giving the orbit space $({\mathbb S}^{(5|4)})^3//S_3\times \mathbb{Z}_2$ 
obtained by dividing the action of $S_3\times\mathbb{Z}_2$ 
on the three copies, where $\mathbb{Z}_2$ is the simultaneous reflection.
This phase corresponds to the Higgs phase of the theory because the gauge symmetry completely breaks.

If one looks into $r\rightarrow 0$ limit, on top of the above Higgs phase, one can have another possibility as it is explained in \cite{OV} and \cite{BV}. 
In this phase, the $\sigma_R^S$ is unconstrained but the matter variables are constrained to satisfy 
\begin{equation}\label{C1}
{\mathcal{O}}_R^S={\bar\phi}_x^S  \phi^x_R +{\bar\phi}_A^S  \phi^A_R=0 
\end{equation} 
The mass term for the fermions and bosons are written as 
\begin{eqnarray}
 {\bar\phi}_x^S \{\sigma,\bar\sigma \}^R_S \phi_R^x+{\bar\phi}_A^S \{\sigma,\bar\sigma \}^R_S \phi_R^A
\end{eqnarray}
And so whenever the $\sigma$ gets expectation value the matter fields become massive and one can integrate them out from the theory. 
One can easily compute the 1-loop correction to the condition (\ref{C1}) which should be proportional to $r$ by doing the path integral with a cut-off $\mu$,
\begin{eqnarray}
\left<{\mathcal{O}}\right>_{\text{1-loop}}&=&-\sum_{A=1}^6 \displaystyle\int d^2 p \frac{1}{p^2+\{\sigma,\bar\sigma\}}+\sum_{x=1}^4 \displaystyle\int d^2 p 
\frac{1}{p^2+\{\sigma,\bar\sigma\}}
\\ \nonumber
&=&-\frac{1}{2\pi}\log \left( \frac{\{\sigma,\bar\sigma\}}{2\mu^2}\right)=r
\end{eqnarray}
which has a solution as
\begin{equation}\label{C2}
 \{\sigma,\bar\sigma\}=2\mu^2 \exp \left(-2\pi r\right)
\end{equation}
After integrating over all the matter fields, the classical vacua $V=0$ is given by condition $\Tr[\sigma,\bar\sigma]^2=0$ which 
together with (\ref{C2}) gives the following solution,
\begin{equation}
 \sigma=\sigma_0 \mu \exp \left( -2\pi r\right)
\end{equation}
where here $\sigma_0$ is an orthogonal $6\times 6$ constant matrix. 
This means that $\sigma$ can be diagonalized and for each diagonal component of the $\sigma$ 
in the small radius regime, one gets a copy of the ${\mathbb S}^{(5|4)}$ as it was seen before. 

\section{Principal chiral model}

In this section, we derive the PCM (principal chiral model) for $Osp(6|4)$. We analyze the differences. 
The model is based on gauging the coset ${Osp(6|4) \over SO(6) \times Sp(4)}$. This is a purely Grassmannian coset manifold with 24 fermionic 
coordinates $\Theta^x_A$. There are other gaugings leading to ${Osp(6|4) \over  Osp(4|2) \times Osp(2|2)}$ and to ${Osp(6|4) \over Osp(4|2) \times SU(1|1,1)}$, 
but we do not discuss them in the present paper. 
Notice that unlike ${Osp(6|4) \over SO(6) \times Sp(4)}$, which has 24 fermions, the 
other two spaces have 12 bosons and 12 fermions. 

The worldsheet action is 
\begin{equation}
S = r^2_{AdS} \int d^2z \, {\rm Str} \, \left(( g^{-1} \p g - A) ( g^{-1} \bar\p g - \bar A) \right)  
\end{equation}

where the indices $x,y$ are raised and lowered with $\e^{xy}$. 
It is invariant under the local symmetry $Osp(6|4)$  under the transformations 
\begin{equation}
\delta g = g \, \Omega\,, 
\delta A = d \Omega + [A, \Omega]\,, 
\end{equation}
where $\Omega \in Osp(6|4)$. We can gauge-fix  the subgroup $SO(6) \times Sp(4)$ by choosing the gauge 
$g = G(\theta, \hat\theta) = {\rm exp}( \theta^x_I Q_x^I +  \theta^{x I} Q_{x I})$. Furthermore, we can gauge-fix the rest of the symmetries by 
choosing the gauge 
\begin{equation}
A^x_I = 0\,,   \bar A_x^I = 0\,.
\end{equation}

This second gauge fixing requires the ghost fields $(\overline Z^x_I, Z_x^I)$ and the antighosts 
$(\overline Y_x^I, Y^x_I)$ with the action 
\begin{equation}
S_{ghost} = r^2_{AdS}\int d^2z \, \Big[ -  Y^x_I (\overline\nabla Z)_x^I + \overline Y_x^I (\nabla \overline Z)^x_I  \Big]\,,
\end{equation}

where 
\begin{equation}(\overline\nabla Z)_x^I = \bar\p Z_x^I + \overline A^{~y}_{x} Z_y^I + \overline A^{I}_{~J} Z_x^J\,, ~~~ 
(\nabla \overline Z)_{I}^x = \p \overline Z_{I}^x +  A^{x}_{~y} \overline Z_{I}^y + A_{I}^{~J} \overline Z_{J}^x \,.
\end{equation}

Assuming that the kinetic term for the remaining gauge fields $A^I_x, \bar A^x_I$ vanishes in the limit 
of large RR fluxes, we can integrate out these fields leading to  get the complete action 
\begin{eqnarray}
S &=& r^2_{AdS} \int d^2z \, \Big[( G^{-1} \p G )^x_I ( G^{-1} \bar\p G)^{I}_x -  
Y^x_I (\overline\nabla Z)_x^I + \overline Y_x^I (\nabla \overline Z)^x_I  
\\ \nonumber
&+&  ( G^{-1} \p G - A)_{xy} ( G^{-1} \bar\p G - \bar A)^{xy} + ( G^{-1} \p G - A)_{IJ} ( G^{-1} \bar\p G - \bar A)^{IJ} 
\\ \nonumber
&+&   ( G^{-1} \p G - A)_{I}^{~J} ( G^{-1} \bar\p G - \bar A)_{J}^{~I} + ( G^{-1} \p G - A)^{IJ} ( G^{-1} \bar\p G - \bar A)_{IJ}
 \Big]\,.
\end{eqnarray}
Notice that the action has the gauge symmetry $SO(6) \times Sp(4)$.  
Eliminating the gauge fields  $A^{xy}\,, \dots\,, \overline A^{IJ}$, one gets a non-linear sigma model which corresponds to the pure spinor 
sigma model with the addition of a BRST exact term (\ref{amodel}).

\section{D-branes and gauge theories}

In order to discuss open strings and D-branes we have to see 
how to put the boundary conditions. 
We start from the supercoset $Osp(6|4)/ SO(6) \times  Sp(4)$. We 
reduce it as follows: the bosonic subcoset: $SO(6) \times Sp(4)$ is 
reduced to $U(3)\times Sp(2)$ and the fermionic part is halved. This achieved by
 using the boundary conditions 
\be
\Theta^{\a I} = \delta_{\dot\a}^{\a} \, {\cal J}^I_J \, \bar\Theta^{\dot\a J}\,, 
~~~~~
\bar\Theta_{I}^{ \dot \a} = \delta^{\dot\a}_\a \, {\cal J}_I^J \, \Theta_{J}^{ \a}\,, 
\ee
where ${\cal J}^I_j$ is the complex structure on ${\mathbb P}^3$. The tensor $\delta_\a^{\dot\a}$ 
reduce the subgroup $Sp(4)$ to $Sp(2)$. We recall that using the symplectic matrices 
$\Lambda$ of $Sp(4, {\mathbb R})$ as the $4\times 4$ matrices satisfying $\Lambda^T \e \Lambda = \e$ 
where $\e = {\rm i} \, \sigma_2 \otimes \ma$, we can see 
immediately the two subgroups $Sp(2, {\mathbb R}) \times Sp(2, {\mathbb R})$. 
In the above equation, we have selected the diagonal subgroup $Sp(2, {\mathbb R})$. 
The above equations are invariant under $Sp(2, {\mathbb R}) \times U(3)$. Notice that we have identified on the boundary of the Riemann surface the 
fermionic variables of the subset ${\cal H}_1 = 
\{\Theta^{\a I}, \Theta_{I}^{\dot\a}\}$ with those of the other subset ${\cal H}_3 = 
\{{\bar\Theta}_{I}^{ \a}, {\bar\Theta}^{\dot\a I}\}$. This simply reduces the 24 fermions to 12 ones. 
The new set of states can be represented in terms 
of the supercoset 
\be
{SU(3|1,1) \over  U(3) \times SU(1,1)}
\ee 
(where we have used the isomorphism $Sp(2, {\mathbb R}) \simeq SL(2,{\mathbb R}) \simeq SU(1,1)$). 
The 6 fermions  are in the $(3, 2)$ or in the $(\bar 3, 2)$ representation of the bosonic 
subgroup.  

In addition, we have to recall $SL(2, {\mathbb R})  \simeq AdS_3$, which can be seen 
by parameterizing a group element of $SL(2,{\mathbb R})$ as follows 
 \be
 g = \left(\begin{matrix} X_{-1} + X_1 & X_0 -X_2 \\ - X_0 -	X_2 & X_{-1} - X_1 \end{matrix}\right)
 \ee
with the condition $\det{g} = X_{-1}^2 - X^2_1 + X_0^2 - X^2_2 = 1$ which shows that the 
$SL(2, \mathbb R)$ group manifold is a 3-dimensional hyperboloid. The metric on $AdS_3$ 
is given by $ds^2 = - d X_{-1}^2 + d X_{1}^2 - d X_{0}^2 + d X_{2}^2$, which is the invariant metric on the group manifold. Then, we have that these boundary 
conditions imply a boundary theory of the type $N=6$ super-YM/Chern-Simons 
on $AdS_3$ space. 

There is another possibility which is given by the following boundary conditions 
\be
\Theta^{\a I} = \delta^\a_{\dot\a} \, \delta^I_J \, \bar\Theta^{\dot\a J}\,, 
~~~~~
\bar\Theta_{I}^{ \dot \a} = \delta_\a^{\dot\a} \, \delta_I^J \, \Theta_{J}^{ \a}\,, 
\ee
In this case the supergroup $Osp(6|4)$ is broken to $Osp(6|2) \times SO(2)$. Notice that 
using the delta $\delta_I^J$ in place of ${\cal J}_I^J$ we do not break the $SO(6)$. In addition, the 
subgroup $Sp(4)$ is broken to $Sp(2) \times SO(2)$. Now, using the isomorphism $SU(4) \simeq SO(6)$, 
we can see the coset $SO(6) \times SO(2) / SU(3) \times U(1) \simeq {\mathbb S}^7/{\mathbb Z}_p$ where $p$ defines how the $U(1)$ is embedded in the groups of the numerator. This observation would help us to lift the 
D-branes solution to KK  monopoles of M-theory. The fermions are halved by the boundary conditions. So, the 
boundary open topological model can be described as the Grassmannian 
\be
\frac{Osp(6|2) \times SO(2) }{U(4) \times Sp(2)}\,.
\ee 
This solution deserves more attention and the study will be postponed in future publications. 

\section{Further directions}

There are several open questions to answer in the framework of gauge/string correspondence
and in particular for this peculiar case given by $AdS_4\times \mathbb{CP}^3$. Here we list 
some of them and we hope to report on them in the near future. 

To complete the program presented here, one needs to explore the cohomology of the BRST 
operator in order to check if the bulk and and the boundary theory describe at least at the linearized level the supergravity states we expect. 
In addition, using the analysis performed in 
\cite{Berkovits:2006ik}, it should be possible to devise a way to define a pure spinor measure for 
tree level and higher loop computations. 
Once this has been established, one of the problems is to prescribe quantum amplitudes for the pure spinor superstring which could be compared 
with super Chern-Simons amplitudes. It would be interesting to single out a subclass of BPS protected 
amplitudes whose string counterpart is therefore calculable via the 
point particle limit and first quantized Chern-Simons theory. 

Having noticed that the vacuum of the target space theory has a Coulomb branch and the relation with the supersphere $\mathbb{S}^{(5|4)}$, 
one is tempted to put the gauge amplitude in relation with a 
topological/twistor string theory on that superspace similarly to \cite{Witten:2003nn}. 

Regarding the boundary field theory, we recall that, using the oscillator technique, the UIR of ${\mathrm {Osp}}(6|4)$ 
are decomposed into representations of its maximal subgroup $SU(3|1,1)$ 
\cite{sing}. 
The singleton is generated out of the vacuum $| 0 \rangle$ and its superpartner $K^{I\alpha} |0\rangle$ where $K^{I \alpha}$ is a 
fermionic oscillator 
in the fundamental representation of $SU(3) \times SU(2)$. The quantum numbers of the vacuum are 
\begin{eqnarray}
|0\rangle &=& | j_0 = 0\,, Q_2 =1 \,, \underline{1} \,, Q_3 = -2 \rangle\,, 
\\ \nonumber
K^{i \a}  |0\rangle &=& | j_0 = 1/2 \,,  Q_2= 2  \,, \underline{3}\,, Q_3 = -1 \rangle\,, 
\end{eqnarray}
where its energy is given by $E_0 = Q_2/2$. These are the only two states annihilated by 
the annihilation operators of the subgroup $SU(3|1,1)$. Acting repeatedly with a 
single-oscillator creation operator ($\a^I$) of $U(3)$ denoted by 
$L^+ = \a^{[I} \a^{J]}$ we get the states 
\begin{eqnarray}
|0\rangle \,, ~~  L^+ |0\rangle  &\longrightarrow& 
\underline{1}(-2) \oplus \underline{3}^*(0)
\\ \nonumber
K^{i \a}  |0\rangle \,, ~~ L^+ K^{i \a}  |0\rangle  &\longrightarrow&
\underline{3}(-1) \oplus \underline{1}(+1)
\end{eqnarray}
The first set is a scalar multiplet that can be recast into a spinorial representation of $SO(6)$, namely 
the fundamental rep $\underline{4}$  of $SU(4)$.   The second set of states forms a multiplet of 
spin $1/2$ fermions in the $\underline{4}^*$ rep of $SU(4)$. 
The number of fields coincides exactly with the content of $D2$ brane counting. 
So, it would be interesting to study the relation between the supersingleton representation and the dual theory 
\cite{DallAgataWZ}.

Of course the relation with M-theory and the membrane theory should be explored also 
in the pure spinor context.

\vspace{1 cm}
{\bf Acknowledgements}
We thank N. Berkovits, P. Fr\'e, A. Tanzini, M. Trigiante and R. D'Auria for very useful discussions. 
H.S. is grateful to DISTA where part of this work was done. 
This research has been supported by the Italian MIUR
under the program ``Teoria dei Campi, Superstringhe e Gravit\`a''.
The work of G.B. and of P.A.G. is supported by
the European Commission RTN Program MRTN-CT-2004-005104.

\end{document}